\documentclass[aps,prd,amsmath,amsfonts,amssymb,eqsecnum,nofootinbib,secnumarabic,superscriptaddress,notitlepage,twocolumn]{revtex4}
\usepackage{graphicx,multirow,slashed,subfigure,xcolor,hyperref}
\usepackage[utf8]{inputenc}
\usepackage[scientific-notation=false,round-mode=figures,round-precision=2]{siunitx}

\providecommand{\tabularnewline}{\\}
\def\pTVeto{p_{T}^{\rm Veto}}
\def\HTVeto{H_{T}^{\rm Veto}}
\def\fb{{\rm ~fb}}
\def\ab{{\rm ~ab}}
\newcommand{\invfb}{{\rm ~fb^{-1}}}
\def\GeV{{\rm ~GeV}}
\def\TeV{{\rm ~TeV}}
\renewcommand{\arraystretch}{1.2}
\newcommand{\be}{\begin{equation}}
\newcommand{\ee}{\end{equation}}
\def\bsp#1\esp{\begin{split}#1\end{split}}

\definecolor{darkgreen}{rgb}{0.0, 0.2, 0.13}

\begin{document}
\title{Sleptons without Hadrons}

\author{Benjamin Fuks}
\email{fuks@lpthe.jussieu.fr}
\affiliation{Laboratoire de Physique Th\'eorique et Hautes Energies (LPTHE),
  UMR 7589, Sorbonne Universit\'e et CNRS, 4 place Jussieu, 75252 Paris Cedex 05, France}
\affiliation{Institut Universitaire de France, 103 boulevard Saint-Michel,  75005 Paris, France}

\author{Karl Nordstr\"om}
\email{knordstrom@lpthe.jussieu.fr}
\affiliation{Laboratoire de Physique Th\'eorique et Hautes Energies (LPTHE),
  UMR 7589, Sorbonne Universit\'e et CNRS, 4 place Jussieu, 75252 Paris Cedex 05, France}
\affiliation{National Institute for Subatomic Physics (NIKHEF) Science Park 105, 1098 XG Amsterdam, Netherlands}

\author{Richard Ruiz}
\email{richard.ruiz@uclouvain.be}
\affiliation{Centre for Cosmology, Particle Physics and Phenomenology {\rm (CP3)},\\
  Universit\'e Catholique de Louvain, Chemin du Cyclotron, 1348 Louvain la Neuve, Belgium}

\author{Sophie L. Williamson}
\email{sophie.williamson@kit.edu}
\affiliation{Laboratoire de Physique Th\'eorique et Hautes Energies (LPTHE),
  UMR 7589, Sorbonne Universit\'e et CNRS, 4 place Jussieu, 75252 Paris Cedex 05, France}
\affiliation{Institute for Theoretical Physics, Karlsruhe Institute of Technology,
Wolfgang-Gaede-Str. 1, 76131 Karlsruhe, Germany.}

\vspace{10pt}
\begin{abstract}
Multilepton searches for electroweakino and slepton  pair production at hadron
colliders remain some of the best means to test weak-scale supersymmetry.
Searches at the CERN Large Hadron Collider, however, are limited by large
diboson and top quark pair backgrounds,
despite the
application of traditional, central jet vetoes. In this context, we report the
impact of introducing dynamic jet vetoes in searches for colorless superpartners.
As a representative scenario, we consider the
Drell-Yan production
of a pair of right-handed smuons decaying into a dimuon system accompanied with
missing transverse energy.
As an exploratory step, we consider several global and local measures of the
leptonic and hadronic activity to construct the veto.
In most all cases, we find that employing a dynamic jet veto
improves the sensitivity, independently of the integrated luminosity.
The inclusion of non-perturbative multiple particle interactions and
next-to-leading order jet merging does not alter this picture.
Directions for further improvements are discussed.
\end{abstract}

\maketitle

\section{Introduction}\label{sec:Intro}

Weak-scale supersymmetry, if realized in nature,
presents an attractive solution to several longstanding theoretical and observational shortcomings of the Standard Model of particle physics (SM).
For example, supersymmetry can protect
the Higgs boson mass from large quantum corrections, ensure gauge
coupling unification at high scales, and provide a viable weakly
interacting dark matter candidate~\cite{Nilles:1983ge,Haber:1984rc}.
While light, sub-TeV  superpartners of quarks and gluons have largely been
excluded by direct searches at the CERN Large Hadron Collider (LHC)~\cite{Aaboud:2017bac,Aaboud:2017vwy,Sirunyan:2017kqq,Sirunyan:2018vjp}, the situation
is far less conclusive for electroweak (EW) boson and lepton superpartners due to
their smaller production cross sections~\cite{Fuks:2012qx,Fuks:2013lya}. Current
constraints only exclude slepton masses up to a few hundreds of GeV~\cite{Sirunyan:2018nwe,Aaboud:2018jiw}. For electroweak boson partners~\cite{Aaboud:2018zeb,Aaboud:2018sua,Sirunyan:2018ubx}, the case is slightly more
interesting due to several small excesses, which reveal a local significance of
$3.5\sigma$ and favor $100-300$ GeV neutralino and chargino masses in the
Minimal Supersymmetric Standard Model (MSSM)~\cite{Athron:2018vxy}. Hence,
studies into new analysis strategies that can improve searches for
electroweakinos and sleptons are highly motivated.

Among the several promising lines of such investigations are those that consider the
impact of jet vetoes ({\it i.e.}, the rejection of events featuring jets with a
transverse momentum greater than some threshold $p_T^{\rm Veto}$~\cite{Barger:1990py,Barger:1991ar,Bjorken:1992er,Fletcher:1993ij,Barger:1994zq}) in measurements of and
searches for heavy, colorless SM~\cite{Banfi:2012jm,Becher:2013xia,Stewart:2013faa,Meade:2014fca,Jaiswal:2014yba,Monni:2014zra,Campanario:2014lza,Gangal:2014qda,Becher:2014aya,Gangal:2016kuo,Michel:2018hui,Stewart:2010pd,Berger:2010xi,Becher:2012qa,Jager:2018cyo} and beyond the SM~\cite{Baer:1993ew,Baer:1995va,Andreev:2006sq,Tackmann:2016jyb,Ebert:2016idf,Fuks:2017vtl,Pascoli:2018rsg,Pascoli:2018heg} states.
Interestingly, recent studies of multilepton searches for heavy, colorless exotic particles have demonstrated that
dynamic jet vetoes can significantly improve discovery potential~\cite{Pascoli:2018rsg,Pascoli:2018heg}.
More specifically, a proposed analysis premised on setting $\pTVeto$ on an event-by-event basis
to the hardness $(p_T)$ of the event's leading lepton was found to improve
sensitivity by roughly an order of magnitude.
The improvement followed from an increase (relative to a static jet veto) in signal rate passing the jet veto,
an ability to veto top quark events without heavy quark flavor-tagging,
and a sensitivity to jets misidentified as charged leptons~\cite{Pascoli:2018rsg}.
While serving a similar goal, such a veto functions in a qualitatively
different manner than rapidity-dependent vetoes~\cite{Gangal:2014qda,Gangal:2016kuo,Michel:2018hui} by
associating $\pTVeto$ with a measure of the hard process scale $Q$.
For $WH/WZ$ production, 
a spiritually similar veto definition using the transverse energy $(E_T)$ of final-state weak bosons and jets was 
proposed in the parton-level study of Ref.~\cite{Campanario:2014lza}.
A key point is that the improvement,
which was demonstrated for both the Drell-Yan (DY) and electroweak boson fusion processes,
followed from the veto effectively discriminating local leptonic activity against local hadronic activity~\cite{Pascoli:2018heg}.

\begin{figure}
  \centering
  \includegraphics[width=0.5\textwidth]{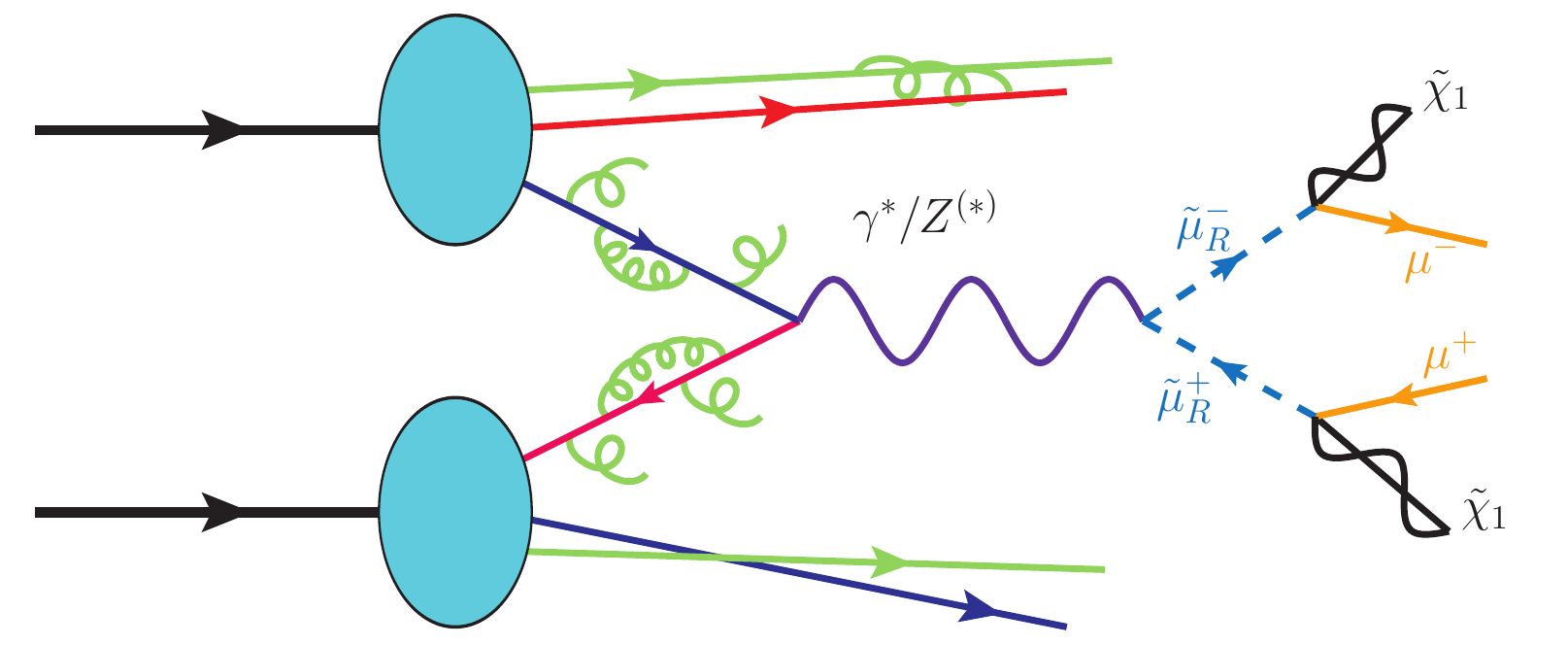}
  \caption{Drell-Yan production of a pair of right-handed smuons ($\tilde{\mu}_R^+
    \tilde{\mu}_R^-$) decaying into a pair of muons ($\mu^\pm$) and lightest neutralinos
    ($\tilde{\chi}_1$). Generated with \textsc{JaxoDraw}~\cite{Binosi:2003yf}.}
  \label{fig:feynman_DYX_SMuonR_MuX}
\end{figure}

In light of this, we have explored the impact of dynamic jet vetoes on the
discovery potential of dimuon plus missing energy searches for right-handed
smuon pairs $(\tilde{\mu}_R^+\tilde{\mu}_R^-)$ decaying to neutralinos $(\tilde{\chi}_1)$ via the DY mode,
\begin{equation}
  q\overline{q} \to \gamma^*/Z^{(*)} \to \tilde{\mu}_R^+\tilde{\mu}_R^-
    \to \mu^+\mu^- \tilde{\chi}_1 \tilde{\chi}_1\ ,
  \label{eq:dySmuons_Parton}
\end{equation}
as illustrated in fig.~\ref{fig:feynman_DYX_SMuonR_MuX}. 
We go beyond Refs.~\cite{Pascoli:2018rsg,Pascoli:2018heg},
which determined only the improved sensitivity of setting $\pTVeto$
to the leading lepton $p_T$, and consider several complementary
measures of local and global leptonic and hadronic activity, including the
scalar sum over lepton transverse momenta ($S_T$) as well as the (inclusive)
scalar sum over the transverse momenta of hadronic objects ($H_T$).
As a benchmark, we use a CMS-inspired analysis~\cite{Sirunyan:2018nwe} that features a
standard (flavor-independent), static, central jet veto of $\pTVeto=25\GeV$.
As will be shown below, a dynamic veto can improve the
discovery potential of the analysis in most cases.
We also explore briefly the impact of including non-perturbative, multiple
particle interactions as well as next-to-leading-order (NLO) jet merging, and find little impact on our conclusions.
For the former, this agrees with previous reports on static jet vetoes~\cite{Jager:2018cyo}.

The remainder of this report continues in the following manner: in
sec.~\ref{sec:theory}, we introduce our simplified model describing right-handed
smuon production and decay in hadron collisions, and discuss the present
constraints on the model. In sec.~\ref{sec:mc}, we summarize our
computational setup, which includes state-of-the-art event
generation up to NLO in QCD matched to parton showers
(PS). We discuss slepton pair production at the LHC and the qualitative impact
of different dynamic jet vetoes for the signal and background processes in
sec.~\ref{sec:signalProc}. There we also define our proposed dynamic veto and
benchmark collider analyses. In sec.~\ref{sec:results} we present our results
and outlook, before summarizing and concluding in sec.~\ref{sec:conclusions}.

\section{Model}\label{sec:theory}

In order to investigate smuon production in a model-independent way, we focus on
a benchmark simplified model inspired by the MSSM. We consider an MSSM limit in
which all superpartners are decoupled, with the exception
of the right-handed smuon $\tilde{\mu}_R$ (of mass $m_{\tilde{\mu}_R}$) and the
lightest neutralino $\tilde{\chi}_1$ (of mass $m_{\tilde{\chi}_1}$) that is taken as
bino-like. The Lagrangian describing the new physics dynamics of our model is
given, using four-component fermion notations, by
\be\bsp
 & \mathcal{L}  =
   \big[\partial_\mu\tilde{\mu}_R^\dag\big]\big[\partial^\mu\tilde{\mu}_R\big]
    \!+\! \frac{i}{2} \overline{\tilde{\chi}_1}\slashed{\partial}\tilde{\chi}_1
    \!-\! m_{\tilde{\mu}_R}^2 \tilde{\mu}_R^\dag\tilde{\mu}_R
    \!-\! \frac12  m_{\tilde{\chi}_1}  \overline{\tilde{\chi}_1}\tilde{\chi}_1\\
 & \qquad
    + \bigg[\partial^\mu \tilde{\mu}_R^\dag \tilde{\mu}_R -
        \tilde{\mu}_R^\dag \partial^\mu \tilde{\mu}_R\bigg]
      \bigg[ i e A_\mu - \frac{i e s_W}{c_W} Z_\mu\bigg]\\
  &\qquad
    -  \frac{\sqrt{2} e}{c_W} \bigg[ \Big(\overline{\tilde{\chi}_1} P_R \mu\Big)
        \tilde\mu_R^\dag  + {\rm H.c.} \bigg].
 \label{eq:lagrangian}
\esp\ee
Here, we have explicitly indicated the smuon gauge interactions with the photon
$A_\mu$ and $Z$ boson field $Z_\mu$ (first term of the second line), as well as the supersymmetric
gauge interactions of the muon $\mu$, the smuon $\tilde\mu_R$, and the bino
$\tilde{\chi}_1$ (last term of the second line). As irrelevant for our purposes, $D$-term contributions
are  neglected. In our notation, $s_W$ and $c_W$ are the sine
and cosine of the electroweak mixing angle, $e$ is the electromagnetic
coupling constant, and $P_R$ the right-handed chirality projector.

Despite its simplicity, the model is only weakly constrained by LHC searches for smuon
pair production in the dimuon plus missing transverse energy channel~\cite{Sirunyan:2018nwe}.
This is due to large backgrounds, consisting mainly of $W$ boson and top quark\
pair production,
as well as being an electroweak signal production mode, as illustrated by eq.~\eqref{eq:dySmuons_Parton}.
For a massless neutralino, the
smuon mass is constrained with $\mathcal{L}=39.5\invfb$ of $\sqrt{s}=13\TeV$ data
to satisfy, at the 95\% confidence level (CL),
$m_{\tilde{\mu}_R}  > 220\GeV$.
There is almost no constraint when the neutralino is heavier than 100\GeV.

As the neutralino is stable, it is a viable candidate for a dark matter
particle. Bino dark matter with light sleptons can be accommodated provided that the
slepton-neutralino mass splitting
is of at most 10\% of the neutralino
mass. Under this condition, there is sufficient co-annihilations so that the
universe is not over closed~\cite{Baker:2018uox}. However, in the aim of using
simplified models as tools for characterizing given phenomena, this latter
constraint is ignored.

\section{Computational Setup}\label{sec:mc}

To conduct our study, we simulate and analyze signal and background events in
$pp$ collisions at a center-of-mass energy $\sqrt{s}=14\TeV$. We implement the simplified model Lagrangian of
eq.~\eqref{eq:lagrangian} into \textsc{FeynRules}~\cite{Alloul:2013bka}, that we
jointly use with the NLOCT~\cite{Degrande:2014vpa} and {\sc FeynArts}~\cite{Hahn:2000kx} packages to generate a UFO library~\cite{Degrande:2011ua} that
includes tree-level vertices as well as ultraviolet and $R_2$ counterterms.
This enables numerical computations up to one-loop in the strong coupling constant $\alpha_s$. 
Event generation for signal and background processes is
performed with \textsc{MadGraph5\_aMC@NLO}~v2.6.3.2~\cite{Alwall:2014hca}, allowing us to match NLO QCD fixed-order calculations with
parton showers with the MC@NLO prescription~\cite{Frixione:2002ik}.
For background samples, the totally inclusive process at NLO in QCD is matched to its first jet multiplicity at NLO according to the FxFx method~\cite{Frederix:2012ps}.
This has the effect of promoting the first and second QCD emissions in the inclusive sample, which, respectively, are only described at LO+LL and LL precision,
to NLO+LL and LO+LL quantities.
In these instances, the generator-level cuts $p_T^j>30\GeV$ and $\vert \eta^j \vert < 5$ are applied with a merging scale $Q_{\rm cut} = 60\GeV$.
We use the {\sc MadSpin}~\cite{Artoisenet:2012st} and {\sc MadWidth}~\cite{Alwall:2014bza} programs
to handle the smuon decays into a muon--neutralino system.

We use \textsc{Pythia} v8.230~\cite{Sjostrand:2014zea},
steered by the CUETP8M1 ``Monash*'' tune~\cite{Skands:2014pea},
to handle parton showering (including QED radiation), the hadronization of all
final-state partons, as well as the decays of hadrons and tau leptons.
Background processes are dressed with multiple particle interactions (MPI) using {\sc Pythia}~8's underlying event model~\cite{Sjostrand:1987su,Sjostrand:2004pf,Sjostrand:2014zea}.
To account for recoil against dipole radiation in the parton shower and for color reconnection between the hard scattering system and proton beams,
we further tune {\sc Pythia}~8 with:
\begin{verbatim}
    SpaceShower:dipoleRecoil=on
    TimeShower:globalRecoil=off
    ColourReconnection:mode=1
    BeamRemnants:remnantMode=1
\end{verbatim}
Particle-level reconstruction is handled with \textsc{MadAnalysis5} v1.7.10~\cite{Conte:2012fm,Conte:2018vmg},
in which we enforce jet clustering using the anti-$k_T$
algorithm~\cite{Cacciari:2008gp}, as implemented in
\textsc{FastJet}~v3.3.0~\cite{Cacciari:2011ma}.
We choose a jet radius of $R=1$, following the jet veto analysis of
ref.~\cite{Fuks:2017vtl}. During the clustering procedure, ideal $b$-jet,
light-jet, and hadronic tau ($\tau_h$) tagging is assumed; 
momentum smearing due to mismeasurement and 
potential misidentification of one particle species as another is implemented at the analysis
level as done in Ref.~\cite{Pascoli:2018heg}. 
Computations use the \textsc{NNPDF 3.1 NLO+LUXqed} parton distribution
function (PDF) set~\cite{Bertone:2017bme}, while both PDF and $\alpha_s(\mu)$
evolutions are managed by using LHAPDF 6~v1.7~\cite{Buckley:2014ana}.

{
In the above Monte Carlo setup, we do not explicitly simulate pileup, 
{\it i.e.}, multiple, simultaneously occurring and spatially overlapping $pp$ collisions.
While this is an important experimental issue for the HL-LHC, the success of
state-of-the-art pileup mitigation techniques
({\it e.g.}, Refs.~\cite{Cacciari:2007fd,ATLAS:2014cva,CMS:2014ata,Bertolini:2014bba})
greatly ameliorates contamination by minimum bias events~\cite{ATLAS:2014cva,CMS:2014ata,Soyez:2018opl}.
A net impact of these pileup-subtraction methods is usually the additional smearing of a particle's four-momentum, 
which is captured by our momentum smearing procedure.
We therefore ignore the presence of pile-up jets in individual events.
While non-negligible, their impact is expected to be subdominant compared to events with high jet multiplicity and underlying events, 
which are taken into account.
}

In addition to event generation, totally inclusive cross section normalizations
at NLO and with next-to-leading logarithmic (NLL) threshold
corrections are obtained with \textsc{Resummino}~v2.0.1~\cite{Fuks:2013vua}. We use again the \textsc{NNPDF 3.1 NLO+LUXqed} PDF set,
despite the availability of PDFs extracted using threshold-corrected matrix
elements~\cite{Bonvini:2015ira}. Our choice is motivated by the much
larger statistical uncertainty of the resummed PDF, which
obfuscates their improved perturbative precision / systematic uncertainty. We
refer to ref.~ \cite{Fiaschi:2018xdm} for a study of their impact on
the hadroproduction of slepton pairs.

For signal rate normalization up to NLO+ NLL(threshold), the collinear factorization $(\mu_f)$ and
QCD renormalization $(\mu_r)$ scales are set to the smuon mass.
For signal and background event generation, we set scales on an event-by-event basis
to half the scalar sum of the transverse energy of all final-state particles,
\begin{equation}
   \mu_{f,r} = \xi \times \mu_0, \text{with}\quad
   \mu_0 = \frac12 \sum_{k\in\{\text{final~state}\}}
       \sqrt{\vert p_T^k\vert^2 + m^2_k}.
\end{equation}
The parton shower scale $(\mu_s)$ is set dynamically to~\cite{Alwall:2014hca}
\begin{equation}
\mu_s = \xi \times \tilde{\mu}_0,  \quad\text{with}\quad \tilde{\mu}_0 \approx \min\left[\mu_0, \sqrt{d_*}\right],
\end{equation}
where $d_* = \min\{d_i\}$ is the minimum $k_T$-distance measure~\cite{Catani:1993hr,Ellis:1993tq} over all QCD parton splittings in the hard process.
By default, we set $\xi=1$. The residual perturbative scale dependency is then
quantified by varying $\mu_f,~\mu_r$, and $\mu_s$, independently over the discrete range
$\xi \in \{0.5, 1.0, 2.0\}$.

\section{Smuon Pairs at the LHC}\label{sec:signalProc}

\subsection{Smuon Pair Production}

Like electroweakinos, sleptons can be produced through a variety of mechanisms
in proton-proton collisions. For simplicity, we restrict ourselves to
right-handed smuon pair production through the inclusive, Drell-Yan process,
\begin{equation}
pp \to \gamma^*/Z^* +X\to \tilde{\mu}_R^+\tilde{\mu}_R^- +X,
\label{eq:dySmuons_Hadron}
\end{equation}
as illustrated in fig.~\ref{fig:feynman_DYX_SMuonR_MuX}.
At the hadronic level $X$ above denotes an arbitrary number of (predominantly
forward) QCD jets. If vector boson fusion becomes a relevant production mode of
TeV-scale smuons~\cite{Gunion:1987ir,Cho:2006sx,Konar:2006qx}, as for example at
higher collider energies and integrated luminosities beyond the LHC, then one
can expect much of the same dynamic jet veto behavior as presented below~\cite{Pascoli:2018heg}.

\begin{figure}
  \centering
  \includegraphics[width=\columnwidth]{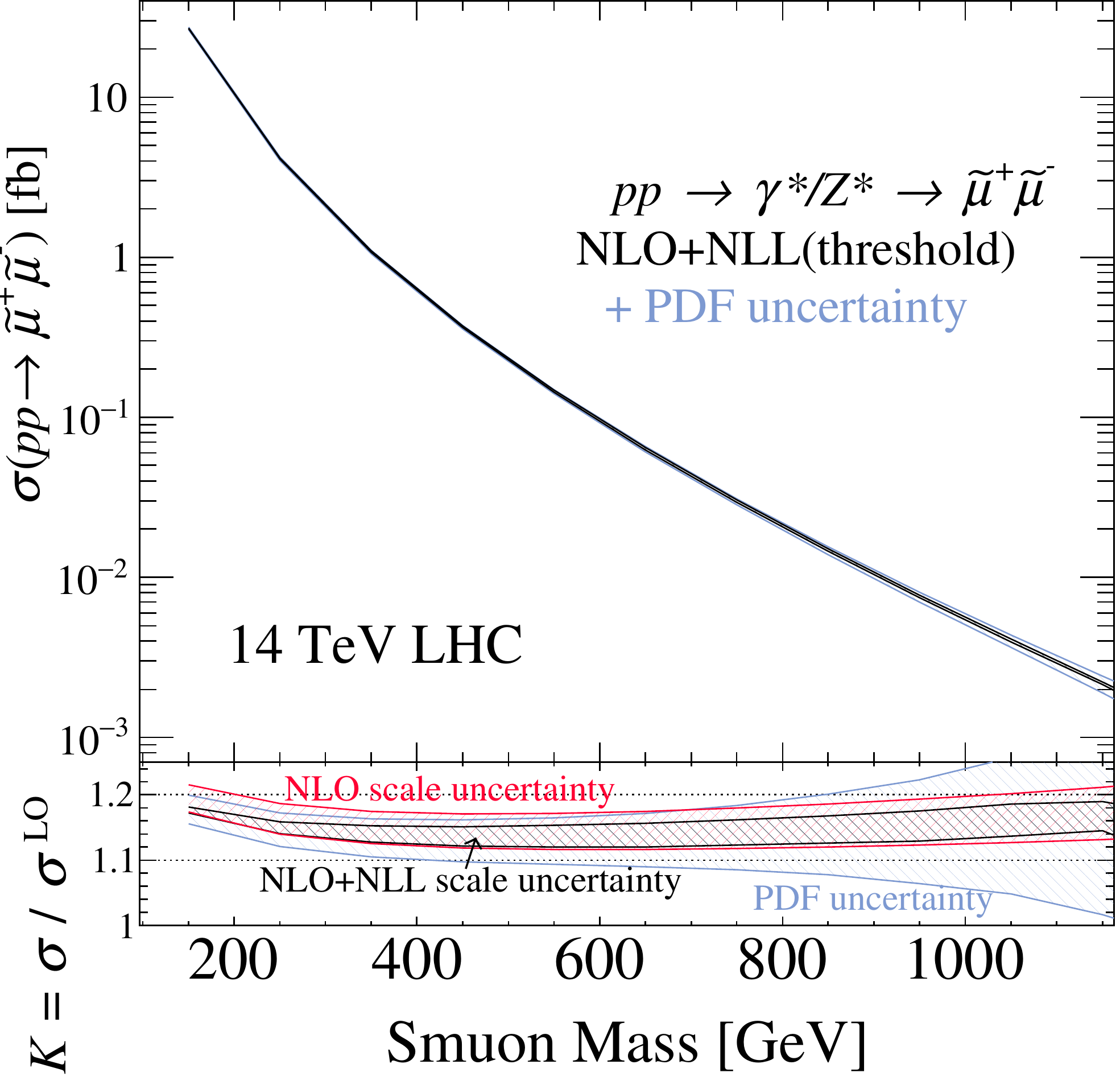}
  \caption{Upper: totally inclusive neutral-current DY production cross section of
  $\tilde{\mu}_R^+\tilde{\mu}_R^-$ pairs at NLO+NLL, and at a center-of-mass
  energy $\sqrt{s}=14\TeV$ with scale uncertainty (black band) and PDF uncertainty (lightest band).
Lower: NLO+NLL (black band) and NLO (lighter band) QCD $K$-factor, with PDF uncertainty (lightest band).
  }
  \label{fig:sleptonVeto_XSec_vs_mSusy_LHCX14}
\end{figure}

In the upper panel of fig.~\ref{fig:sleptonVeto_XSec_vs_mSusy_LHCX14}, we show
the totally inclusive NLO+NLL cross section for neutral-current DY smuon
production at a center-of-mass energy $\sqrt{s}=14\TeV$. The results are given
as a function of the smuon mass, and we indicate
the uncertainties stemming from perturbative scale variation (black band)
and PDF fitting (light band). In the lower panel of the figure, we present QCD
$K$-factors, with their uncertainties, defined relative to the Born process,
\begin{equation}
K^{\rm NLO+N^kLL} = \frac{\sigma^{\rm NLO+N^kLL}(pp \to \tilde{\mu}_R^+\tilde{\mu}_R^- +X)}{\sigma^{\rm LO}(pp\to \tilde{\mu}_R^+\tilde{\mu}_R^- +X)}.
\end{equation}
The cases $k<0$ and $k=1$, respectively, correspond to computations at NLO and NLO+NLL(threshold).

For smuon masses $m_{\tilde{\mu}_R} \in [200, 900]\GeV$ ({\it i.e.}, the
range of interest for the LHC), the NLO+NLL production cross section varies from
approximately 10\fb~to 10\ab, with the corresponding scale uncertainties
reaching the $\pm2-3\%$ level. In this mass regime, NLO+NLL
predictions sit within the NLO perturbative uncertainty band that has a
width of about $\pm4\%$. Furthermore, the QCD $K$-factors for both the
NLO and NLO+NLL computations are about $K\approx 1.15$ and largely
independently of the smuon mass. On different grounds and still in this mass
range, PDF uncertainties are only marginally larger than the NLO scale
uncertainties, before growing significantly for $m_{\tilde{\mu}_R}\gtrsim 800\GeV$
(due to the absence of data in the PDF fits).
As the same PDF set is used for both the NLO and NLO+NLL computations,
the size of their uncertainties is essentially identical.

For the parameter space consistent with our simplified model assumptions, the
gluon fusion contribution to inclusive $\tilde{\mu}_R^+\tilde{\mu}_R^-$
production, which formally arises at $\mathcal{O}(\alpha_s^2)$, is small
compared with the neutral-current DY component~\cite{Borzumati:2009zx,Lindert:2011td}.
Moreover, for DY-like processes that give rise to high-$p_T$ charged leptons,
QCD scale uncertainties in cross sections featuring a dynamic jet veto at NLO+PS
(which are formally at the leading-logarithmic accuracy) are comparable with the total
inclusive cross section uncertainty at NLO due to the absence of large jet veto logarithms~\cite{Pascoli:2018rsg,Pascoli:2018heg}.
This holds independently of the jet radius for a dynamic veto~\cite{Pascoli:2018heg}.
 For a static veto, choosing a jet radius of
 $R=1$ greatly helps to minimize the perturbative uncertainties~\cite{Dasgupta:2014yra,Banfi:2015pju,Dasgupta:2016bnd,Fuks:2017vtl},
though worsens the universal, non-perturbative ones~\cite{Becher:2014aya,Fuks:2017vtl,Dasgupta:2007wa}.
Thus, we may conclude that cross sections for $\tilde{\mu}_R^+\tilde{\mu}_R^-$ production obtained from event generation at NLO+PS,
either with or without a dynamic jet veto, are reliable estimates of the true rate.
Similar reliability of NLO+PS predictions with static jet vetoes applied to SM diboson and weak boson scattering processes have been reported elsewhere~\cite{Monni:2014zra,Jager:2018cyo}.
Hence, for our purposes and for discovery purposes,
NNLO and NNLL(threshold) terms in fixed order and resummed signal predictions can be ignored.

\subsection{Dynamic Jet Vetoes Beyond $p_T$}\label{sec:signalProc_beyondPT}

Jet vetoes have long been established as powerful tools to improve the discovery potential of sleptons and electroweakinos in multilepton searches
at hadron colliders~\cite{Baer:1993ew,Baer:1995va,Andreev:2006sq,Aaltonen:2008pv,Tackmann:2016jyb}.
In practice, LHC experiments rely on fixed/static veto thresholds of $\pTVeto=20-50\GeV$
for central jets within a pseudorapidity $\vert \eta^j \vert \lesssim 2.5$~\cite{Aad:2012pxa,Aad:2014vma,Khachatryan:2014qwa,Aad:2014yka,Aad:2015eda,Sirunyan:2017lae,Aaboud:2018jiw}.
Recently~\cite{Pascoli:2018rsg,Pascoli:2018heg}, though, it was demonstrated that dynamic jet veto schemes,
namely ones wherein $\pTVeto$ is set on an event-by-event basis to the $p_T$ of an event's  leading lepton,
can improve the sensitivity of multilepton searches for exotic, colorless particles.
In conjunction with selection cuts on leptonic observables, this type of jet veto ultimately discriminates against the relative amounts of  hadronic and leptonic activity in each event.

In this sense, dynamic jet vetoes can be generalized by considering observables that measure an event's global hadronic and leptonic activities instead of just the $p_T$ of an event's leading objects.
Natural candidates include: 
the inclusive scalar sum of $p_T$ of all hadron clusters in an event $(H_T^{\rm Incl.} )$,
\begin{equation}
H_T^{\rm Incl.} \equiv \sum_{k\in\{{\rm clusters}\}} \vert \vec{p}_T^{~k} \vert, \quad \vert \eta^{k}\vert \lesssim 4.5,
\label{eq:defHT}
\end{equation}
or  the exclusive scalar sum of $p_T$ of the two leading charged leptons $(\ell_1, \ell_2)$ in an event $(S_T^{\rm Excl.} )$,
\begin{equation}
S_T^{\rm Excl.} \equiv \sum_{k=1}^2 \vert \vec{p}_T^{~\ell_k} \vert.
\end{equation}
Here we adopt the usual particle ordering, where $p_T^{k_i} > p_T^{k_{i+1}}$ for particles $k_i$ and $k_{i+1}$ of species $k$.
We also henceforth suppress the ``Incl./Excl.'' labels for brevity but stress that we do not expect results
here to uniformly carry over to exclusive $H_T$ and inclusive $S_T$.

To be explicit, the summation over ``hadron clusters'' in Eq.~(\ref{eq:defHT}) means the summation over the set of momentum vectors 
that are the output of a jet clustering algorithm applied to hadrons within a pseudorapidity of $\vert \eta^{\rm Had.}\vert < \eta^{\max} = 4.5$.
Clusters that satisfy additional kinematic requirements, {\it e.g.}, a minimal $p_T$ threshold, are further classified as jets.
Working with clusters has the highly nontrivial impact of reducing the size (and complexity) of individual events,
and therefore Monte Carlo data sets, as well as ameliorating some (but not all) of the impact of underlying event (UE) / multiple particle interactions (MPI).
This follows from the fact that, 
despite the redistribution of hadron multiplicity and average hadron $p_T$ induced by UE/MPI models,
the spatial distribution of hadrons is always strongly correlated with the hard scattering process itself,
and therefore will be largely recaptured by $k_T$-style sequential clustering algorithms.
Intuitively, one aspect of UE/MPI models is to ``fatten'' clusters by increasing the average number of hadrons in a cluster 
(while decreasing the average $p_T$ per hadron~\cite{Sjostrand:1987su,Sjostrand:2004pf}),
but only alter the kinematics of a cluster of radius $R=1$ and transverse momentum $p_T\gtrsim 25\GeV$
by $\mathcal{O}(\Lambda_{\rm NP}/p_T R)\lesssim5-15\%$, for a non-perturbative scale $\Lambda_{\rm NP}=2-3\GeV$.
Such an estimate is consistent with the findings of dedicated studies on the impact of MPI on jet observables~\cite{Jager:2018cyo,MPIveto}.
Moreover, these shifts impact charged leptons at a comparable level by momentum recoil/conservation.
Taking the ratio of leptonic and hadronic observables, as one does for dynamic jet vetoes, thereby further mitigates the impact of MPI.
We have checked this explicitly for our high-statistics, FxFx-matched background samples and observe that
MPI induces differences in lepton-to-hadron ratios on the order of their perturbative QCD uncertainties, {\it i.e.}, $1-10\%$,
but reach as large as $15\%$ as one nears phase space boundaries, for cluster/jet radii of $R=0.4-1.0$.
As such, to highlight the qualitative behavior of perturbative matrix elements, for the remainder of this section we restrict ourselves to
results obtained with inclusive, NLO+PS samples and postpone further discussion of non-perturbative uncertainties to table~\ref{tab:CMSCutflows}.

Qualitatively, $H_T$ differs from the $p_T$ of the leading (or subleading) central jet  $p_T^{j_1}$ (or $p_T^{j_2}$)
in that $H_T$ is much more sensitive to complicated color topologies in a hard scattering processes.
The simplest color topologies, {\it e.g.}, eq.~\ref{eq:dySmuons_Hadron},
have at most one or two color dipoles / antennas, and hence less QCD radiation, resulting in $H_T$ that is comparable to $p_T^{j_1}$.
On the other hand, complex QCD processes, {\it e.g.}, $pp \to WW+nj$, have many color antennas, and hence more sources of QCD radiation, resulting in $H_T$ significantly larger than $p_T^{j_1}$.
Metaphorically speaking, $H_T$ {\it vs.} $p_T^{j_1}$ is like a multiband {\it vs.} single-band radio emitter,
with complex color structures inducing many bands of radiation simultaneously whereas signal-like topologies have fewer bands.

\begin{figure*}
\centering
\subfigure[]{\includegraphics[width=.48\textwidth]{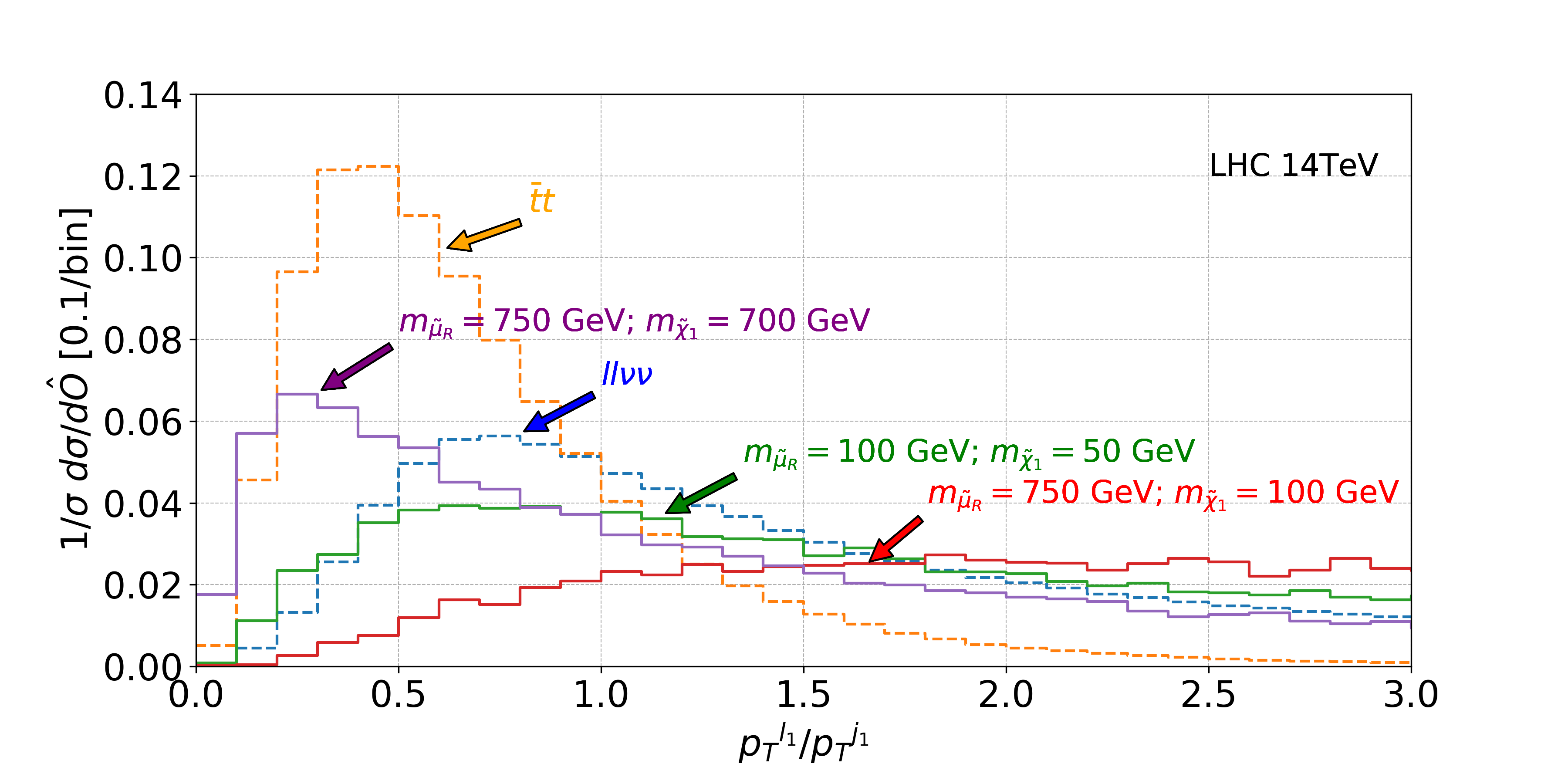}	}
\subfigure[]{\includegraphics[width=.48\textwidth]{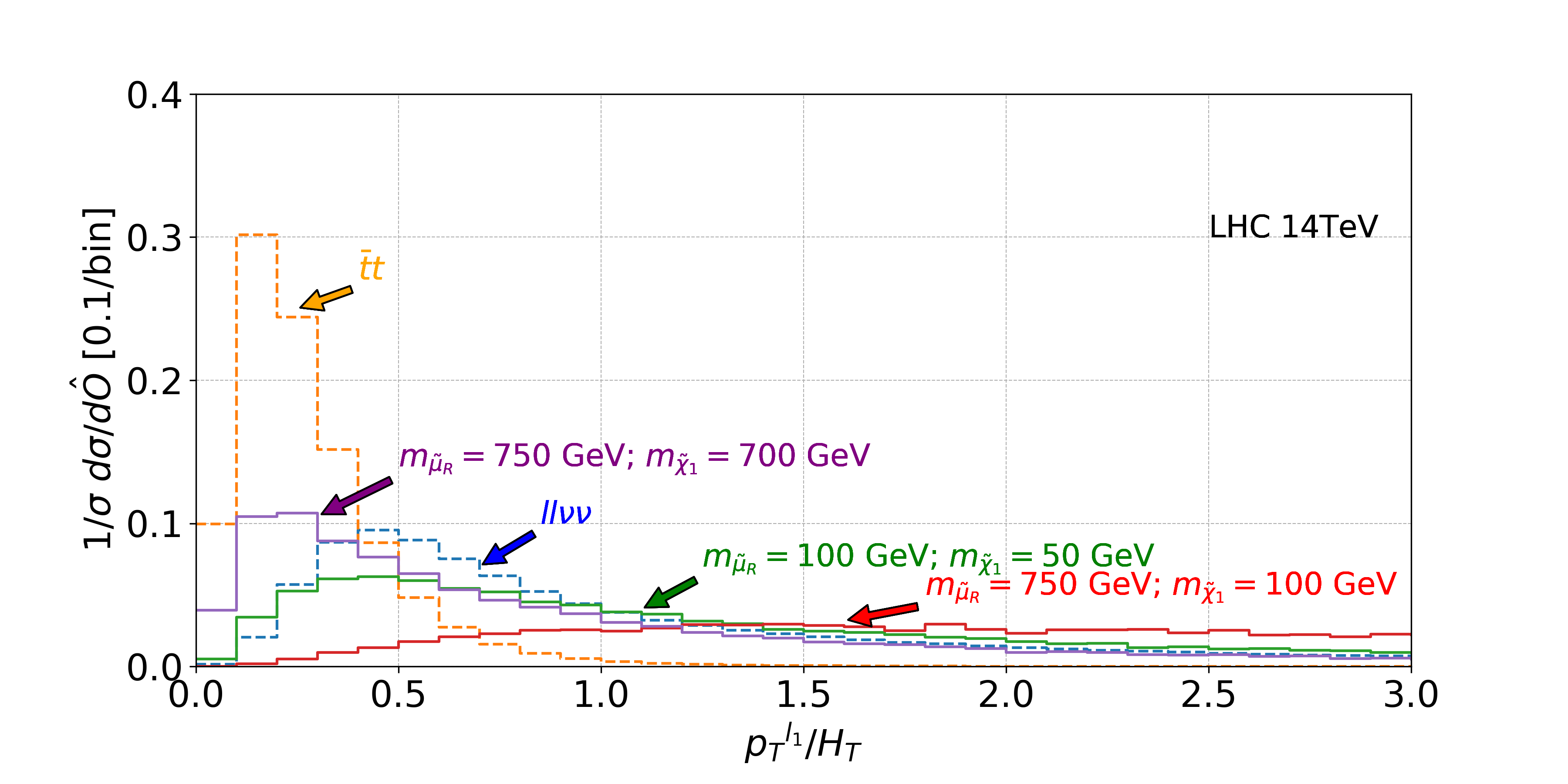}	}
\\
\subfigure[]{\includegraphics[width=.48\textwidth]{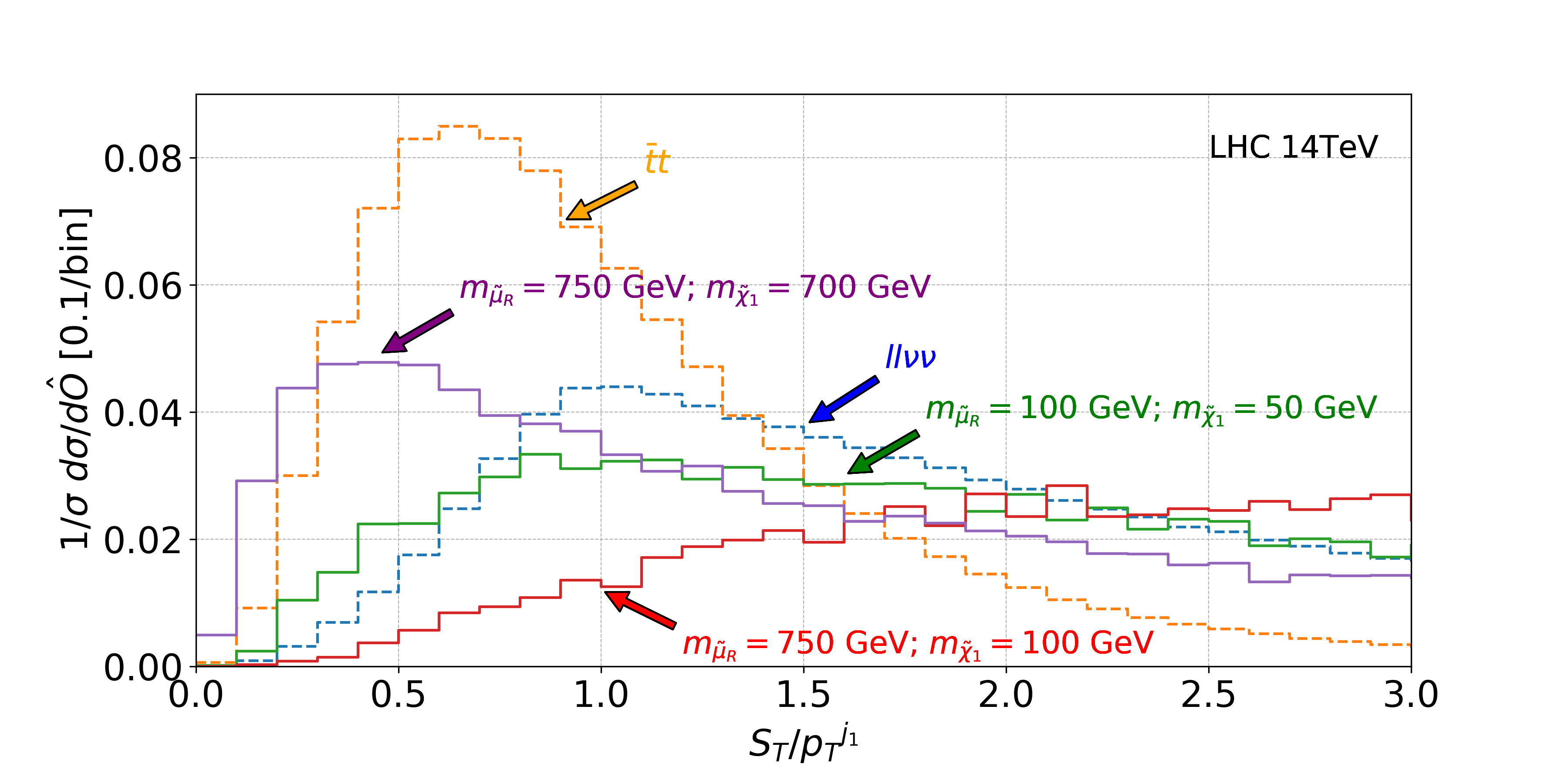}}
\subfigure[]{\includegraphics[width=.48\textwidth]{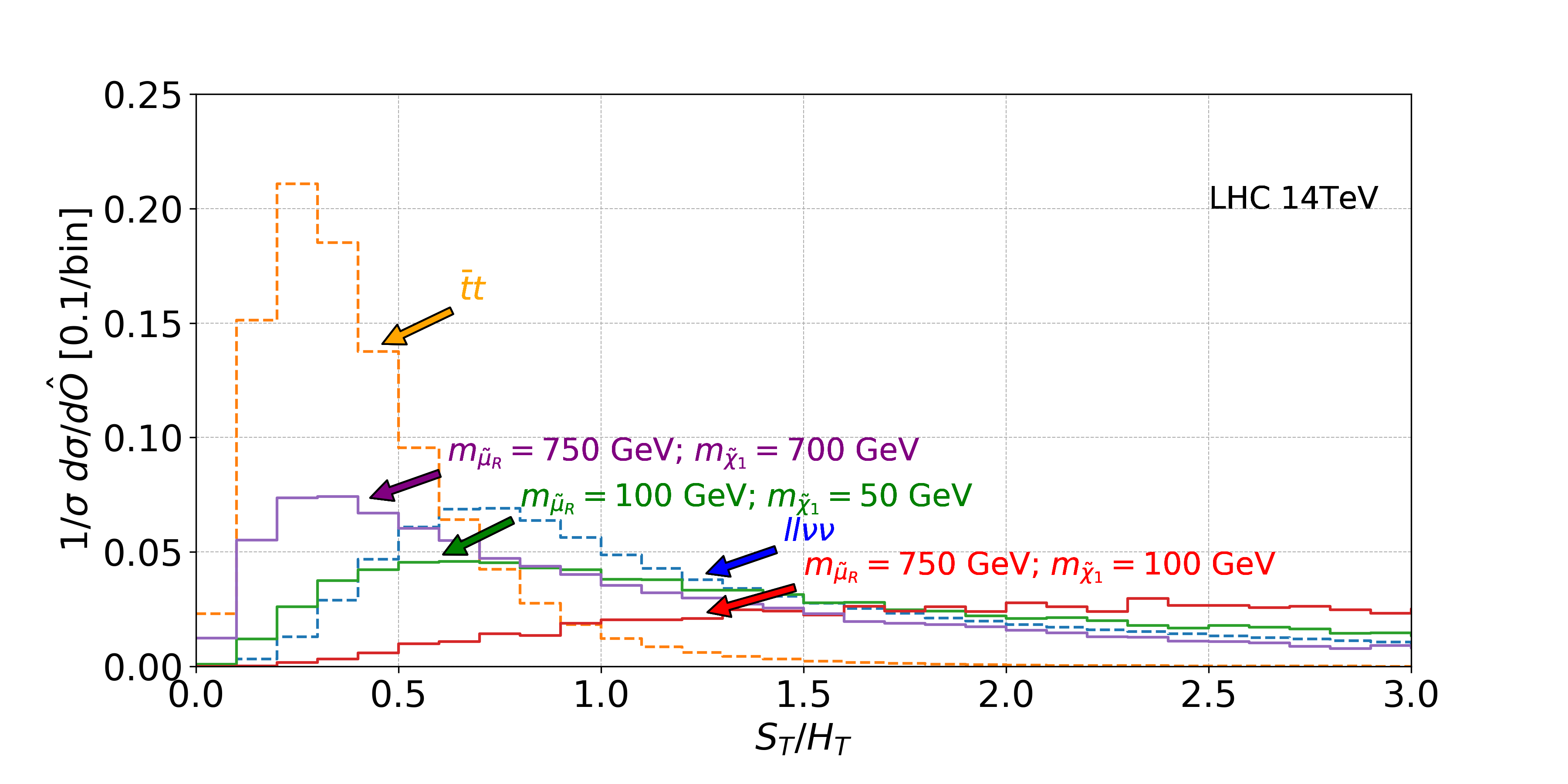}	}
\\
\subfigure[]{\includegraphics[width=.48\textwidth]{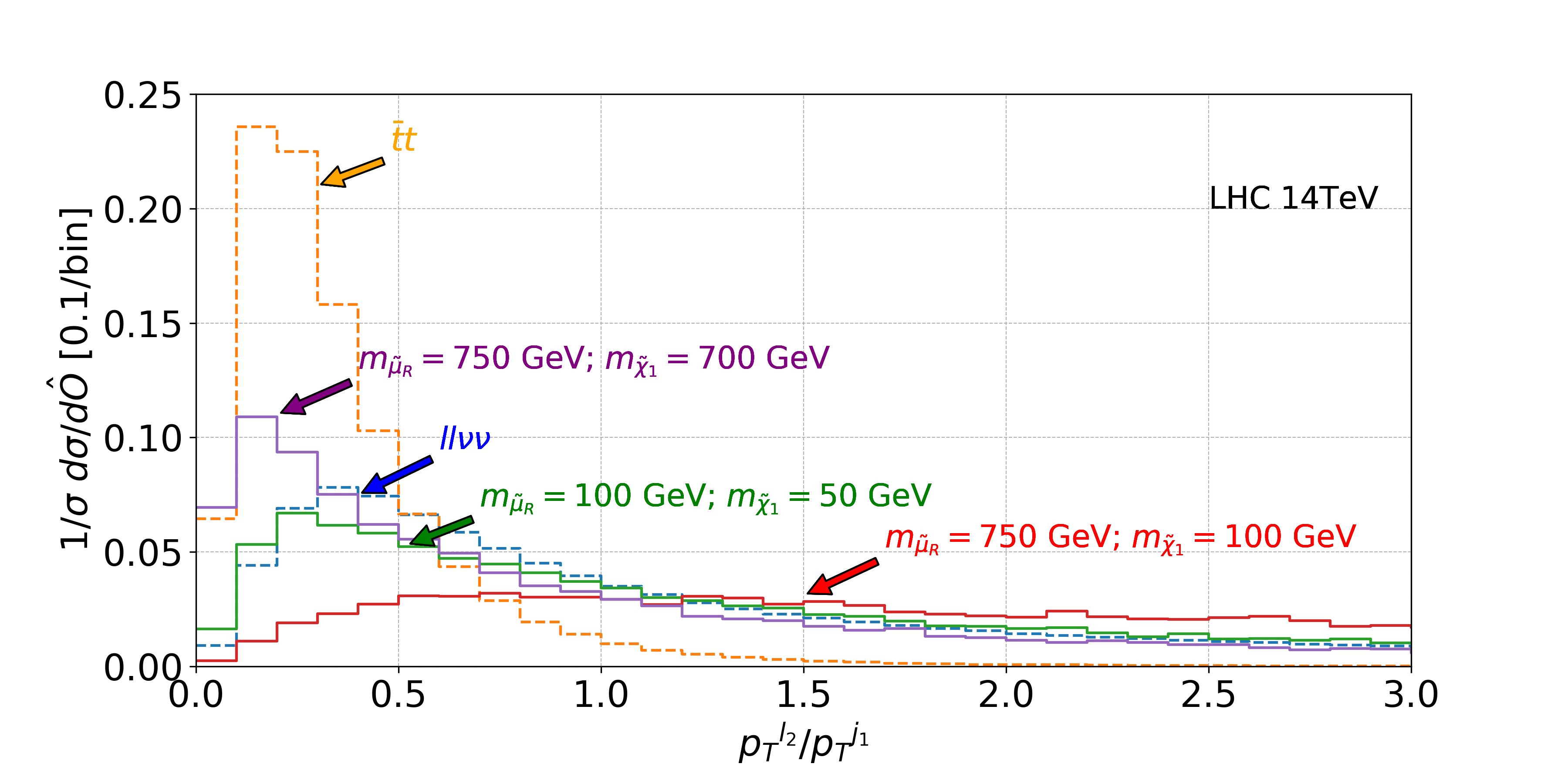}}
\subfigure[]{\includegraphics[width=.48\textwidth]{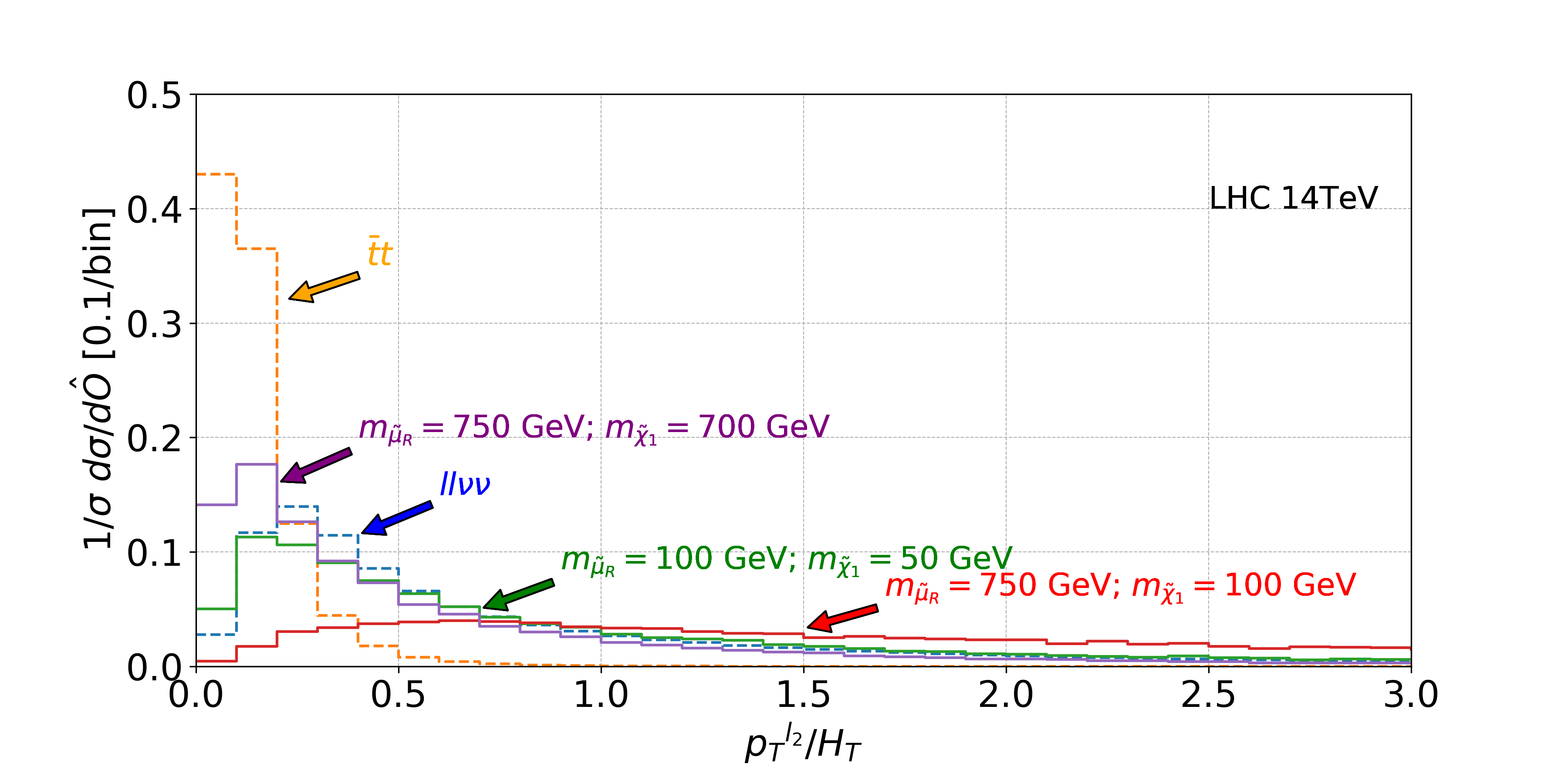}}
\caption{Ratios of measures of hadronic and leptonic activity for representative signal (solid) and background (dashed) samples used in the dynamic veto analysis, showing 
(a) $p_T^{\ell_1}/p_T^{j_1}$, 
(b) $p_T^{\ell_1}/H_T$, 
(c) $S_T/p_T^{j_1}$, 
(d) $S_T/H_T$, 
(e) $p_T^{\ell_2}/p_T^{j_1}$, 
(f) $p_T^{\ell_2}/H_T$.}
\label{fig:DynamicRatios}
\end{figure*}

Due to its exclusive nature, $S_T$ acts to exaggerate and accentuate the characteristic behavior of the leading charged leptons
$\ell_1$ and $\ell_2$. If they stem from a resonant (continuum) process, then $S_T$ will characteristically have a narrow (shallow) distribution.
If the two are pair-produced, then one expects the scaling $S_T\sim 2 p_T^{\ell_1}$.
Likewise, any relative (in)dependence of $p_T^{\ell_k}$  on the hadronic activity is inherited by $S_T$.
By virtue of the Collinear Factorization Theorem,
central, high-$p_T$ charged leptons in hadron collisions stem from a hard underlying process.
Hence, the $S_T$ of leading leptons probes an event's hard-scattering core, and,
up to possible kinematic decay factors scales like the hard scattering scale $Q$.
This helps to protect against the emergence of large veto logarithms.
We remark briefly that
exclusive $S_T$ differs from inclusive $S_T$ in that the latter sums over the trailing charged leptons and additionally probes universal, low-$Q^2$ physics, such as hadron decays and QED parton showering.

In application, a dynamic, $H_T$-based jet veto would work, for example, by rejecting events in which $H_T$ exceeds $p_T^{\ell_1}$.
Analogously, an $S_T$-based veto functions by requiring, for example, an event to satisfy $p_T^{j_1} < S_T$ for $\vert \eta^{j_1}\vert < \eta^{\rm max}$.

To explore these alternative dynamic veto schemes, we present in fig.~\ref{fig:DynamicRatios},
the normalized distributions for the following ratios of leptonic and hadronic activities:
\begin{align}
&{\rm (a)}~p_T^{\ell_1} / p_T^{j_1},	\quad		{\rm (b)}~p_T^{\ell_1} / H_T,		\quad	{\rm (c)}~S_T / p_T^{j_1}, \nonumber \\
&{\rm (d)}~S_T / H_T,			\quad		{\rm (e)}~p_T^{\ell_2} / p_T^{j_1},	\quad	{\rm (f)}~p_T^{\ell_2} / H_T.		\nonumber
\end{align}
These are considered for the signal process in eq.~\ref{eq:dySmuons_Hadron}, 
with smuons decaying into a SM muon plus a neutralino.
We assume the benchmark parameter space points,
\begin{eqnarray}
\text{Signal Category}					&:& (m_{\tilde{\mu}_R},m_{\tilde{\chi}_1}), \nonumber \\
\text{High-mass, Large mass splitting}  		&:&  (750\GeV, 100\GeV), \nonumber \\
\text{High-mass, Small mass splitting}  		&:&  (750\GeV, 700\GeV), \nonumber \\
\text{Low-mass, Small mass splitting}  		&:&  (100\GeV, 50\GeV). \nonumber
\end{eqnarray}
We also consider the representative backgrounds,
\begin{equation}
 pp \to t \bar{t} \to \ell^+ \ell^- +X, \qquad
 pp \to \ell^+ \ell^- \nu \overline{\nu},
\end{equation}
with $\ell \in \{e,\mu,\tau_h\}$.
All signal and background processes are considered at NLO+PS, after jet clustering.
For the present situation, we require at least two oppositely charged muons with any number of hadronic clusters satisfying the truth-level kinematical requirements
\begin{equation}
\vert \eta^{\rm clust.}\vert<4.5, \quad
\vert \eta^{\ell} \vert < 2.4, \quad\text{and}\quad p_T^{\ell} > 10\GeV.
\end{equation}

As a reference point, we discuss first the kinematic ratio $r^{\ell_1}_{j_1} = p_T^{\ell_1} / p_T^{j_1}$, as studied by refs.~\cite{Pascoli:2018rsg,Pascoli:2018heg} 
and shown in fig.~\ref{fig:DynamicRatios}(a).
For the signal processes, we see a difference in behavior according to whether or not the smuon and neutralino are close in mass.
Whereas the high-mass, large mass splitting configuration possesses a very broad distribution, with most of the phase space exceeding $r^{\ell_1}_{j_1}>1$,
the more compressed configurations possess relatively narrower distributions, with significantly more phase space below the $r^{\ell_1}_{j_1}=1$ threshold.
For the large mass splitting case, final-state muons carry $p_T^{\ell} \sim m_{\tilde{\mu}_R}(1 - m_{\tilde{\chi}_1}^2 / m_{\tilde{\mu}_R}^2)/2 \sim m_{\tilde{\mu}_R}/2\sim 375\GeV$.
 This is significantly larger than the leading jet $p_T$, which is generally of the order of the Sudakov peak.
For on-shell slepton pair production, the Sudakov peak is much lower than $2m_{\tilde{\mu}_R}$, indicating that characteristically $p_T^{j_1} \ll p_T^{\ell_1} \sim m_{\tilde{\mu}_R}/2$.
For the compressed cases, the muons carry only $p_T^{\ell}\lesssim 40-50\GeV$
and drive the relationship $r^{\ell_1}_{j_1}(\text{high-mass,~small-splitting.})\lesssim r^{\ell_1}_{j_1}(\text{low-mass,~small-splitting.})\lesssim 1$.

Considering the background processes, one observes that most events populate the region around $r^{\ell_1}_{j_1}\sim 0.25-0.75$.
In both cases, the behavior follows from kinematic arguments~\cite{Pascoli:2018rsg}.
For an at-rest top quark decaying into leptons, the characteristic momenta of the charged lepton and associated $b$-quark  give rise to the scaling
\begin{equation}
r^{\ell_1}_{j_1} \sim \frac{p_T^\ell}{p_T^b} \sim \cfrac{m_t\left(1+M_W^2/m_t^2\right)/4}{m_t\left(1-M_W^2/m_t^2\right)/2} \sim 0.75.
\end{equation}
In a full simulation at NLO+PS with large-$R$ jets, this is pushed significantly to smaller values due to a large $t\overline{t}+1j$ sub-channel,
boosts from  large $(t \overline{t})$-invariant masses, and into-cone radiation.
Each enhances $p_T^{j}$ or $p_T^b$ relatively to  $p_T^\ell$.
Despite being a color-singlet process, the inclusive $pp\to\ell\ell\nu\nu+X$ channel has a relatively large $pp\to\ell\ell\nu\nu+1j$ fraction.
This is due to the $pp\to W\gamma^*/WZ+0j$ processes being suppressed by radiation amplitude
zeroes~\cite{Mikaelian:1977ux,Brown:1979ux,Mikaelian:1979nr,Zhu:1980sz,Brodsky:1982sh,Brown:1982xx,Baur:1995uv,Gehrmann:2014fva}.
In turn,  $r^{\ell_1}_{j_1}$ is inherently less than unity.

\begin{figure*}
\centering
\subfigure[]{\includegraphics[width=.48\textwidth]{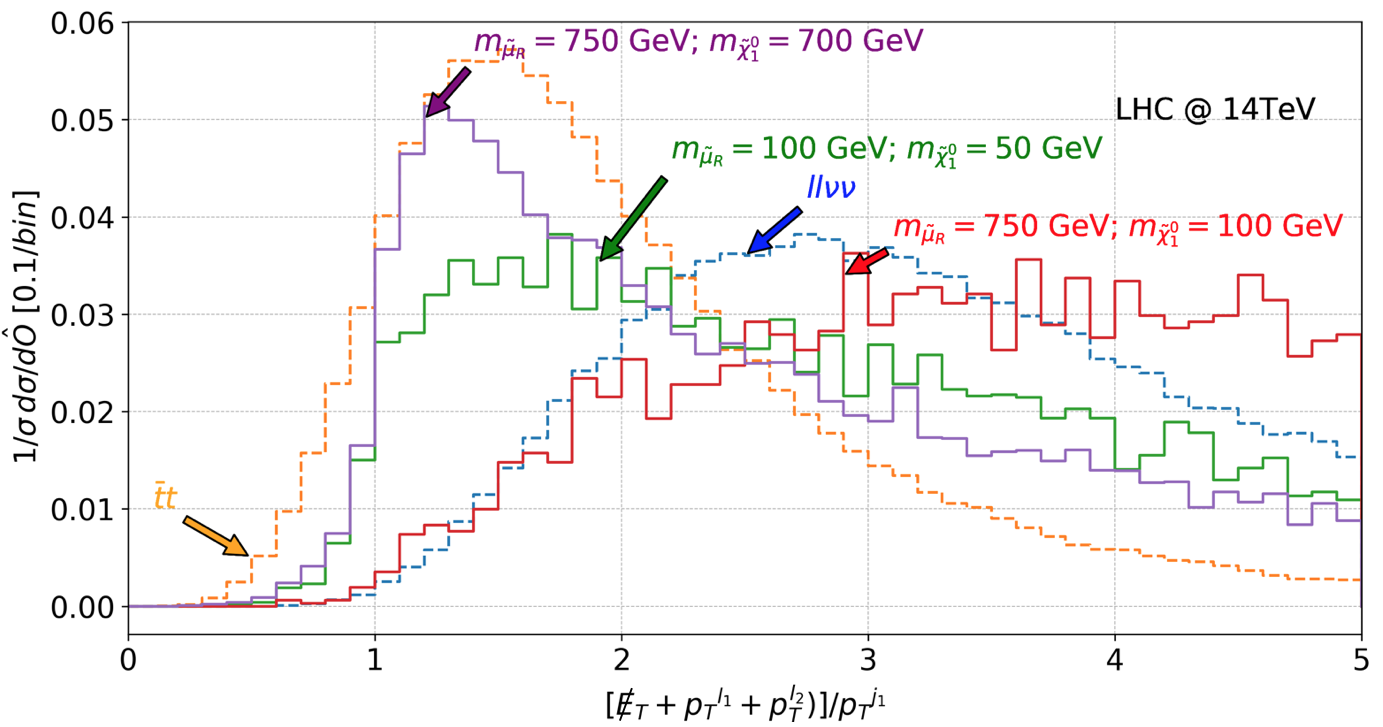}	 }
\subfigure[]{\includegraphics[width=.48\textwidth]{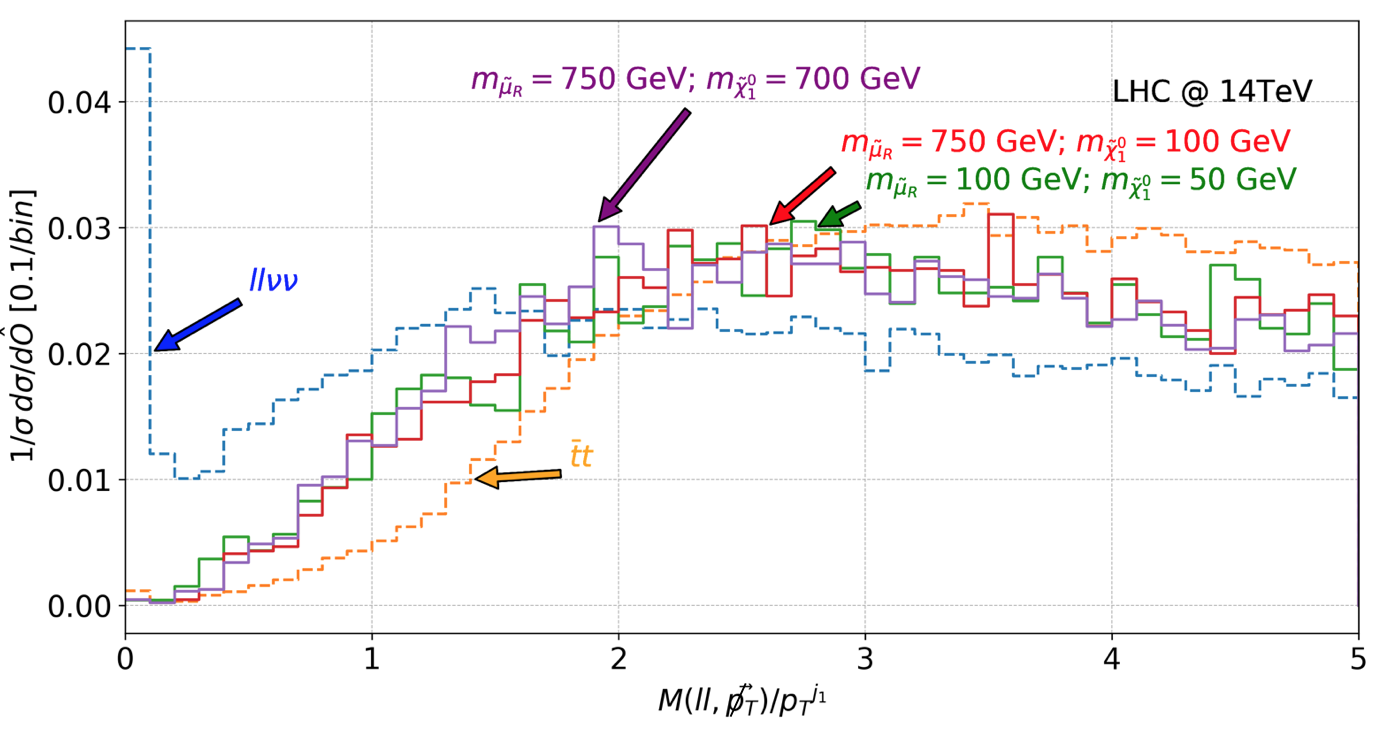}	}
\caption{
{Same as fig.~\ref{fig:DynamicRatios} but for the ratios (a)$ \left(\slashed{E}_T + p_T^{\ell_1} + p_T^{\ell_2} \right)/p_T^{j_1}$ and (b) $M(\ell_1,\ell_2, \slashed{\vec{p}}_T)/p_T^{j_1}$.
}
}
\label{fig:DynamicVetoes_MET}
\end{figure*}

In fig.~\ref{fig:DynamicRatios}(b), we consider the impact of including secondary QCD radiation and show the distribution for  $r^{\ell_1}_{H_T} = p_T^{\ell_1} / H_T$.
For the signal processes, we observe some difference from $r^{\ell_1}_{j_1}$ in the normalization and position of the distributions' maxima.
Here, the maxima are marginally taller and pushed to slightly lower values of  $r^{\ell_1}_{H_T}$.
This is indicative of the low hadronic activity in DY-like processes, which is in fact why a jet veto is considered at all.
On the other hand, for both background processes, we observe values of $r^{\ell_1}_{H_T}$ much smaller than $r^{\ell_1}_{j_1}$.
For $t\overline{t}$ specifically, the shift (and narrowing) from $r^{\ell_1}_{j_1} \lesssim 0.5$ to $r^{\ell_1}_{H_T}\lesssim 0.25$ is consistent with $H_T$, which sums over both bottom jets,
being roughly $1/H_T \sim 1/(2\times p_T^{j_1}) \sim 1/(2\times p_T^{b_1})$.
The low-mass, compressed signal distribution is in particular hardly distinguishable from the $\ell\ell\nu\nu$ distribution.

Considering now a more global measure of leptonic activity, we present in figs.~\ref{fig:DynamicRatios}(c) and (d) the distributions for the ratios
$r^{S_T}_{j_1} = S_T / p_T^{j_1}$ and $r^{S_T}_{H_T} = S_T / H_T$, respectively.
 For all cases we see that the $r^{S_T}_{j_1}$ and  $r^{S_T}_{H_T}$ curves are broader than their $r^{\ell_1}_{j_1}$ and $r^{S_T}_{j_1}$ counterparts,
 and that the distributions' maxima are shifted slightly rightward.
As in the (a) and (b) panels, the compressed signal and both background processes have a significant fraction of their respective phase spaces below unity.

As an alternative measure of local leptonic activity, we show in fig.~\ref{fig:DynamicRatios}(e) and (f) the distributions for the ratios
$r^{\ell_2}_{j_1} = p_T^{\ell_2} / p_T^{j_1}$ and $r^{\ell_2}_{H_T} = p_T^{\ell_2} / H_T$, respectively.
One sees a larger separation than in (a) and (c) of the high-mass, compressed signal process from all other processes.
Notably, the $t\overline{t}$ distributions are much narrower, with almost all events falling below $r^{\ell_2}_{j_1} \lesssim 0.5$ and $r^{\ell_2}_{H_T} \lesssim 0.25$.

Taken together, a picture emerges  for generalized definitions of dynamic jet vetoes.
We find that all of the proposed veto schemes exhibit uniform behavior.
For the signal process with the highest charged lepton momenta, {\it i.e.}, the high-mass, large mass splitting signal category,
we find a clear signal-to-background separation against representative background processes.
For signal processes with charged lepton momenta comparable to SM processes, we find significantly less but nonetheless interesting discriminating power.
In particular, for the low-mass, compressed category, we observe reasonable separation from $t\overline{t}$ but poor separation from $\ell\ell\nu\nu$,
whereas for the high-mass, compressed category we report the opposite.
This suggests that it may be possible to salvage additional signal space with complementary selection cuts.
Quantitatively, we observe a larger signal-to-background separation for dynamic veto schemes with more inclusive/global hadronic observables, {\it e.g.}, $H_T$,
and more exclusive/local charged lepton observables, {\it e.g.}, $p_T^{\ell_2}$.
The worst separation is given by $r^{S_T}_{j_1}$, which makes use of the multilepton activity of background processes but not the relatively low hadronic activity of the signal processes.
The ratio $r^{\ell_2}_{H_T}$ appears to be exceptionally powerful in rejecting top quark background.

{\subsubsection*{Dynamic Jet Vetoes with Missing Transverse Momentum}
A full and systematic investigation of all possible dynamic jet vetoes that one can build for the dimuon final-state 
is beyond our exploratory and proof-of-concept scope.
Nevertheless, it is interesting and constructive to briefly explore the potential of dynamic jet vetoes built from missing transverse momentum observables.
Here and throughout our study, the magnitude ($\slashed{E}_T$) of the transverse momentum imbalance vector ($\slashed{\vec{p}}_T$) 
is defined with respect to all visible momenta within $\vert \eta \vert < 4.5$,
\begin{equation}
\slashed{E}_T = \vert \not\! \vec{p}_T \vert,  \quad \not \!\vec{p}_T = - \sum_{k\in\{{\rm visible}\}} \vec{p}_T^{~k}.
\label{eq:metDef}
\end{equation}
Building observables that include both charged lepton momenta and  $\slashed{\vec{p}}_T$ (or $\slashed{E}_T$) 
takes into account high-energy, invisible particles in the final state ({\it e.g.}, $\tilde{\chi}_1,~\nu$)
and hence are potentially more closely tied to the momentum vector $(\hat{Q})$ and scale $(Q)$ of hard-scattering processes than charged lepton momenta alone.

Assuming the same configurations as for fig.~\ref{fig:DynamicRatios}, we show in fig.~\ref{fig:DynamicVetoes_MET}
the normalized distributions for the ratios
\begin{eqnarray}
&(a)& r_{j_1}^{L_T} =  L_T / p_T^{j_1}, 
\\
&(b)& r_{j_1}^{M(\ell_1, \ell_2, \slashed{\vec{p}}_T)} = M(\ell_1,\ell_2,\not \!\vec{p}_T)/p_T^{j_1}.
\end{eqnarray}
Here, the kinematic observables $L_T$ and $M(\ell_1, \ell_2,  \slashed{\vec{p}}_T)$,
denote, respectively,
($i$) the scalar sum of leading lepton $p_T$, subleading lepton $p_T$, and  $\slashed{E}_T$, as well as 
($ii$) the invariant mass of the dilepton-$\slashed{\vec{p}}_T$ system.
Symbolically, the two are defined by the following:
\begin{eqnarray}
&(i)& L_T =  S_T +  \slashed{E}_T =  p_T^{\ell_1} + p_T^{\ell_2} +  \slashed{E}_T,
\\
&(ii)& M(\ell_1,\ell_2,\not \!\vec{p}_T) = \sqrt{(p_{\ell_1}+p_{\ell_2}+\not \!\vec{p}_T)^2}.
\end{eqnarray}
In fig.~\ref{fig:DynamicVetoes_MET}(a), one sees a strong qualitative resemblance to the ratio $r^{S_T}_{j_1}$ in fig.~\ref{fig:DynamicRatios}(c),
which is expected due to the ratios' similar definitions, but with all curves move rightwards due to the (trivial) inequality $r^{L_T}_{j_1} \geq r^{S_T}_{j_1}$.
Quantitatively, the shift essentially pushes the ratio for each process to values greater than unity with the notable exception of the $t\overline{t}$ sample.
This case features a modest tail to smaller values and can be attributed to uncommon (but not rare) phase space configurations.
Such configurations include for example when the two final-state $b$-jets are both centrally produced (low $\vert\eta\vert$) 
while the $\mu^+\mu^-$ pair are forward with a large $\vert\eta_{\mu_1}-\eta_{\mu_2}\vert$ separation
and the neutrino pair have a maximal azimuthal separation, $\vert\phi_{\nu_1}-\phi_{\nu_2}\vert \sim \pi$.
Here the neutrinos' transverse momentum vectors cancel and 
the charged leptons are sufficiently forward that $p_T^{\ell_1}+p_T^{\ell_2} < p_T^{b_1}$.
As with $r^{S_T}_{j_1}$, the ratio $r_{j_1}^{L_T}$ does very little to separate signal benchmark points from the background, with an
exception for the high-mass, large mass splitting category at large values of $r_{j_1}^{L_T}$.

Figure~\ref{fig:DynamicVetoes_MET}(b) shows a somewhat interesting distribution,
namely a strong separation between the $\ell\ell\nu\nu$ sample from everything else
but relatively poor separation between the smuon and top quark processes.
More specifically, the $\ell\ell\nu\nu$ sample populates $r_{j_1}^{M(\ell_1, \ell_2, \slashed{\vec{p}}_T)} \ll 0.1$,
and can be attributed to the situation where the transverse momentum of the dilepton system, $p_T(\ell_1,\ell_2)$,
strongly recoils against the leading jet in the $\ell\ell\nu\nu+1j$ process.
For such configurations, the leading jet $p_T$ is approximately the recoil of charged leptons and neutrinos.
Hence, the squared invariant mass $M^2(\ell_1,\ell_2,\not \!\vec{p}_T)$ is,
{\small \begin{eqnarray}
M^2(\ell_1,\ell_2,\not \!\vec{p}_T) &=& \left(E^{\ell_1}+E^{\ell_2},~\vec{p}_T^{\ell_1}+\vec{p}_T^{\ell_2}+\not\!\vec{p}_T, p_Z^{\ell_1}+p_Z^{\ell_2}\right)^2
\nonumber\\
&\approx& \left(E^{\ell_1}+E^{\ell_2}, ~-\vec{p}_T^{j_1},  p_Z^{\ell_1}+p_Z^{\ell_2} \right)^2
\nonumber\\
&=& p_T^2(\ell_1,\ell_2) - p_T^{j_1~2},
\end{eqnarray}}%
and reveals a large cancellation for large $p_T^{j_1}$. Subsequently, in this approximation, the ratio is given by
\begin{equation}
r_{j_1}^{M(\ell_1, \ell_2, \slashed{\vec{p}}_T)} \approx \cfrac{\sqrt{p_T^2(\ell_1,\ell_2) - p_T^{j_1~2}}}{p_T^{j_1}},
\end{equation}
and tends toward zero for large $p_T^{j_1}.$
For the signal processes, its Drell-Yan-like topology ensures that the emission of high-$p_T$ jets is disfavored,
leading to larger $r_{j_1}^{M(\ell_1, \ell_2, \slashed{\vec{p}}_T)}$.
For the $t\overline{t}$ process, the leading jet $p_T$ is not a good approximation of 
the sum of leptonic transverse momenta and the $\slashed{\vec{p}}_T$ vector due to the contribution of additional high-$p_T$ jets, particularly a second $b$-jet.
The inclusion of jets into  $\vert\not\!\vec{p}_T\vert$ (and to some extent $p_T^{\ell_k}$) leads to larger values of $M$, and hence larger ratios.
}

\subsection{Jet Veto Collider Analyses}\label{sec:jetVetoAna}
We now turn to defining our static and dynamic jet veto analyses  to quantify how generalized dynamic jet vetoes may improve the discovery potential of smuon pairs at the LHC,
if at all.
For all analyses, we define analysis-quality charged leptons and jets as those that satisfy the following kinematical, fiducial, and isolation requirements:
\begin{eqnarray}
&&\! p_T^{e~(\mu)~[\tau_h]~\{j\}} > 10~(10)~[20]~\{25\}\GeV,   \nonumber\\
&&\! \vert \eta^{e~(\mu)~[\tau_h]~\{j\}} \vert < 2.4, \qquad \!\!\!\! \Delta R_{\ell_m,\ell_n} > 0.4, \quad \Delta R_{\ell j } > 0.4. \qquad \nonumber
\end{eqnarray}
We use the electron and muon efficiencies as reported\footnote{\url{https://twiki.cern.ch/twiki/bin/view/CMSPublic/SUSMoriond2017ObjectsEfficiency}} in ref.~\cite{Sirunyan:2018nwe} for leptons with $p_T \geq 20$ GeV,
and those reported in ref.~\cite{Sirunyan:2018omt} for leptons with $p_T \in [10,20[$ GeV.
We tag the hadronic decays of $\tau$ leptons $(\tau_h)$ with $p_T \geq 20$ GeV using the efficiencies reported in ref.~\cite{CMS-DP-2018-009}.
{To account for finite detector resolution and pileup mitigation techniques,}
all objects are smeared with a Gaussian profile as done in ref.~\cite{Pascoli:2018heg},
with smearing coefficients set using publicly available resolution parametrizations
reported by the ATLAS and CMS collaborations~\cite{CMS:2015kjy,Sirunyan:2017ulk,Chatrchyan:2011ds,Aaboud:2018kfi}.
The missing transverse momentum 2-vector and its magnitude are defined above in Eq.~\ref{eq:metDef}.

We simulate the following background processes,
\begin{eqnarray}
	pp \to \ell\ell\ell\nu,  			&\qquad& 	pp \to \ell\ell\nu\nu,	\nonumber\\
 	pp \to t \bar{t} \to 2\ell X, 		&\qquad&	pp \to WWW \to 3\ell X, \nonumber
\end{eqnarray}
at NLO+PS with FxFx-merging for the first jet multiplicity. The additional background processes
\begin{eqnarray}
 pp \to \ell\ell\ell\ell,  				&\qquad&	pp \to \ell^+\ell^-, 		\nonumber\\
 pp \to t \bar{t} \ell\nu \to 3\ell X, 	&\qquad&  pp \to WW\ell\ell \to 2\ell X, \nonumber
\end{eqnarray}
simulated in contrast at the NLO+PS accuracy, 
were found to give a negligible background contribution after all selection cuts in all analyses,
 and therefore are ignored for the remainder of our report.
{
As our signal process consists of high-$p_T$ muons and large $ \slashed{E}_T$,
QCD multijet backgrounds are similarly negligible~\cite{Sirunyan:2018nwe} 
and ignored for the remainder of this work.}

\subsubsection*{Shared Analysis Baseline}\label{sec:commonAnalysis}

As a baseline for all analyses, we follow closely the CMS search for slepton pair production in dilepton final states at $\sqrt{s}=13$ TeV with $\mathcal{L}=35.9$ fb$^{-1}$ of data~\cite{Sirunyan:2018nwe}.
We pre	select events featuring one pair of analysis quality, opposite-sign muons,
and veto events with additional analysis-quality charged leptons.
We are thus inclusive with respect to additional leptons outside these criteria.
{Such events are identified in LHC collisions through inclusive, low-threshold dimuon triggers.
During Run II, triggers such as the ATLAS experiment's \texttt{2mu10} and \texttt{L1\_2MU10} 
triggers\footnote{See also: \href{https://twiki.cern.ch/twiki/pub/AtlasPublic/TriggerOperationPublicResults/menuTable.png}{\texttt{https://twiki.cern.ch/twiki/pub/AtlasPublic/\\TriggerOperationPublicResults/menuTable.png}}}
require two muon $p_T\gtrsim 10\GeV$ at Level 1~\cite{Aaboud:2016leb}.
It is also possible to identify such events using complementary triggers, such as those used in the aforementioned CMS analysis,
which enable one to lower the threshold for the sub-leading muon to as low as
$p_T>8\GeV$ at the cost of increasing the threshold for the leading muon. By
using multiobject triggers, the CMS collaboration has also managed to push
lower the threshold on
the transverse momenta of the two muons to $p_T\gtrsim3.5$~GeV, at the price of
a moderate missing energy requirement~\cite{Sirunyan:2018iwl}.
}
Low-mass hadronic resonances and $Z$-pole contributions are removed with the invariant mass cuts: $m_{\mu\mu}>20\GeV$ and $\vert m_{\mu\mu} - M_Z\vert > 15\GeV$.
The SM DY continuum is further suppressed by requiring $\slashed{E}_T>100\GeV$, and diboson and top pair processes are reduced by
requiring a ``stransverse mass'' cut of  $M_{T2}>90\GeV$ \cite{Lester:1999tx,Cheng:2008hk}.
In sec.~\ref{sec:results}, we describe the impact of relaxing this cut.
Events are then binned according to $\slashed{E}_T$.
Analysis object definitions  and shared analysis requirements are summarized in the top two sections of table~\ref{tb:SelectionCuts14TeV}.

\subsubsection*{Benchmark, Static Jet Veto Analysis}\label{sec:staticAnalysis}

\begin{table}[t]
\begin{center}
 \begin{tabular}{ c }
\hline \hline
Analysis Object Criteria at $\sqrt{s}=14$ TeV:                                                           \tabularnewline
$p_T^{e~(\mu)~[\tau_h]~\{j\}} > 10~(10)~[20]~\{25\}\GeV,$			 				\tabularnewline
$\vert \eta^{e~(\mu)~[\tau_h]~\{j\}} \vert < 2.4$, \quad anti-$k_T$ w./~$R=1$								 \tabularnewline
$\Delta R_{\ell_m,\ell_n} > 0.4, \quad \Delta R_{\ell j} > 0.4$
 \tabularnewline\hline
Common Analysis Requirements:                                                     				     \tabularnewline
$N(\mu^+) = 1$,  \quad $N(\mu^-) = 1$, \quad $N(\ell)=2$,  				\tabularnewline
$m_{\mu\mu}>20$ GeV, $\vert m_{\mu\mu} - M_Z \vert > 15\GeV$,	\tabularnewline
 $M_{T2}>90\GeV$, ~$\slashed{E}_T>100\GeV$, 						\tabularnewline
 Binned signal region: $\slashed{E}_T\in $~ $(a) [100,150[,$ 				\tabularnewline
$(b) [150,225[, ~ (c) [225,300[, ~(d) [300,\infty[\GeV$
 \tabularnewline \hline
Benchmark (Static) Jet Veto  Analysis Requirements:                                                  	       \tabularnewline
$p_T^{\mu_1~(\mu_2)} > 50~(20)\GeV$,
\quad
$\pTVeto = 25\GeV$
\tabularnewline \hline
Dynamic Jet Veto Analysis Requirements:                                                         			 \tabularnewline
Overlapping Signal Categories: \tabularnewline
(a)~$\pTVeto = p_T^{\ell_1}$ 	\quad  (b)~$\HTVeto = p_T^{\ell_1}$ \tabularnewline
(c)~$\pTVeto = S_T$ 		\quad  (d)~$\HTVeto = S_T$ \tabularnewline
(e)~$\pTVeto = p_T^{\ell_2}$ 	\quad  (f)~$\HTVeto = p_T^{\ell_2}$ \tabularnewline
\hline\hline
\end{tabular}
\caption{
(Top) Analysis object / particle identification requirements at $\sqrt{s}=14\TeV$;
(upper) common analysis requirements;
(lower) benchmark static veto analysis requirements; and
(bottom) dynamic jet veto analysis requirements.
}
\label{tb:SelectionCuts14TeV}
\end{center}
\end{table}

 At this point, our jet veto collider analyses diverge.
 Our benchmark, static jet veto analysis continues as prescribed in the baseline CMS analysis~\cite{Sirunyan:2018nwe}
 and further requires that the $p_T$ of the leading and sub-leading muons satisfy
 \begin{equation}
 p_T^{\ell_1~(\ell_2)} > 50~(20)\GeV.
 \label{cut:lepPT}
 \end{equation}
 Lastly, we impose a static jet veto of $\pTVeto=25\GeV$ on analysis-quality jets.
 As such objects must sit within \mbox{$\vert \eta \vert <2.4$}, the veto is more specifically a static, central jet veto.
 Relaxing this pseudorapidity restriction will be briefly explored in the following section.
Analysis requirements are summarized in the third section of table~\ref{tb:SelectionCuts14TeV}.

For background processes, we find comparable cross sections after selection cuts  to those reported by CMS for all signal regions except the lowest $\slashed{E}_T$ bin.
There, we find that our background rate is about 50\% lower and is driven by a difference in the normalization of the ``Flavor Symmetric'' background,
which is largely populated by the $t\overline{t}$ and diboson processes.
We attribute the difference in this bin to our background normalizations being accurate only up to NLO+PS,
which are therefore missing numerically large $\mathcal{O}(\alpha_s^2)$ contributions, 
and also to potentially missing contributions from mismeasurements which are not captured by our detector fast simulation. 
These effects can both introduce significant differences to CMS' data-driven predictions.

In table~\ref{tab:CMSCutflows}, we present the cut-flows for the $\ell\ell\nu\nu$, ~$\ell\ell\ell\nu$, and $t\overline{t}$ SM backgrounds with their theory uncertainties [$^{+\%}_{-\%}$]  and selection cut efficiencies [(\%)] 
given the analysis cuts in table~\ref{tb:SelectionCuts14TeV}.
Conservative theory uncertainties reported are obtained by adding the renormalization and factorization scale envelope with the shower scale envelope and statistical Monte Carlo uncertainty in quadrature.
For completeness, we explore the backgrounds when modeled at the inclusive NLO+PS level without MPI (Incl.), with MPI (Incl.+MPI), and FxFx-merging with MPI (FxFx+MPI).
Categorically, we observe that the inclusion of MPI does not appreciably impact jet veto cross sections nor their uncertainties for both traditional, static jet vetoes and dynamic jet vetoes.
In particular, we observe changes at the 5-10\% level, inline with uncertainties and findings reported elsewhere~\cite{Jager:2018cyo}.
For the dilepton and trilepton processes, we observe a comparable impact by including FxFx-merging; for $H_T$-based vetoes, the impact of MPI and FxFx slightly compensate for one another.
The theoretical stability afforded by the FxFx sample is consistent with what has been reported elsewhere, {\it e.g.} Refs.~\cite{Jones:2017giv,Chakraborty:2018kqn} and references therein.
For the top quark process, we observe that the impact of both MPI and FxFx are comparable and shift rates in the same direction at a level consistent with uncertainties.
The seemingly qualitative difference from the electroweak cases is due to a sizable increase of the $t\overline{t}$ cross section normalization 
stemming from the virtual correction to the $t\overline{t}j$ subprocess in the FxFx merged sample.
The corresponding virtual correction for the electroweak processes is numerically more modest.
We note that for some top quark cases, {\it i.e.}, the static and dynamic $\HTVeto=p_T^{\ell_2}$ vetoes (see below for the latter), there is a larger statistical Monte Carlo uncertainty driving the total theory uncertainty;
bluntly, these vetoes decimate the several $10^7$-event $t\overline{t}\to 2\ell X$ datasets.

\renewcommand{\arraystretch}{1.8}
\begin{table*}[t]
\setlength\tabcolsep{7pt}
\centering
\resizebox{\textwidth}{!}
{
\begin{tabular}
{>{\centering}m{2.0cm}
|>{\centering\arraybackslash}m{1.5cm}|>{\centering\arraybackslash}m{1.5cm}|>{\centering\arraybackslash}m{1.5cm}
|>{\centering\arraybackslash}m{1.5cm}|>{\centering\arraybackslash}m{1.5cm}|>{\centering\arraybackslash}m{1.5cm}
|>{\centering\arraybackslash}m{1.5cm}|>{\centering\arraybackslash}m{1.5cm}|>{\centering\arraybackslash}m{1.5cm}
}
\multirow{2}{*}{Cut / Channel}
			& \multicolumn{3}{c|}{$\sigma(\ell\ell\nu\nu)$ [fb]} & \multicolumn{3}{c|}{$\sigma(\ell\ell\ell\nu)$ [fb]} & \multicolumn{3}{c}{$\sigma(t\bar{t})$ [fb]} \\
			& Incl.	& Incl.+MPI & FxFx+MPI	& Incl.	& Incl.+MPI & FxFx+MPI	& Incl.	& Incl.+MPI & FxFx+MPI \\
\hline
\multicolumn{10}{c}{Common Analysis Requirements} \\
\hline
Generator	& \num[round-precision=3]{10152}$^{+4.1\%}_{-4.8\%}$	 	& \num[round-precision=3]{10152}$^{+4.1\%}_{-4.8\%}$	& \num[round-precision=3]{10312.0445024}$^{+4.5\%}_{-5.1\%}$
			              & \num[round-precision=3]{1602.1697}$^{+5.5\%}_{-6.5\%}$		& \num[round-precision=3]{1602.1697}$^{+5.5\%}_{-6.5\%}$	& \num[round-precision=3]{1681.8201416}$^{+5.5\%}_{-6.2\%}$
			              & \num[round-precision=3]{85802.18}$^{+9.2\%}_{-10.4\%}$ 		& \num[round-precision=3]{85802.18}$^{+9.2\%}_{-10.4\%}$	& \num[round-precision=3]{91019.26271}$^{+11.9\%}_{-11.7\%}$\\
\hline
Dimuon Selection      	& \shortstack{\\ \num{853.15912}$^{+\num{4.12536095062}\%}_{- \num{4.82896472775 }\%}$ \\ (\num{8.40377770209}$\%)$} 	
					& \shortstack{\\ \num{850.89586}$^{+\num{4.51060821529}\%}_{- \num{5.1326418404 }\%}$ \\ (\num{8.38148409458}$\%)$}	
					& \shortstack{\\ \num{842.95473586}$^{+\num{4.91393819225}\%}_{- \num{5.27626038377 }\%}$ \\ (\num{8.37756465919}$\%)$}
				        &  \shortstack{\\ \num{139.96418}$^{+\num{6.32328228249}\%}_{- \num{6.98932874252 }\%}$ \\ (\num{8.73591696869}$\%)$}	
				        & \shortstack{\\ \num{139.88555}$^{+\num{6.28350445385}\%}_{- \num{6.76263584637 }\%}$ \\ (\num{8.73101077534}$\%)$}	
				        & \shortstack{\\ \num{139.21729964}$^{+\num{6.11110391686}\%}_{- \num{6.51111489733 }\%}$ \\ (\num{8.58401308606}$\%)$}
				        & \shortstack{\\ \num{6445.259}$^{+\num{11.9004300935}\%}_{- \num{11.7032920239 }\%}$ \\ (\num{7.51176906427}$\%)$}		
				        & \shortstack{\\ \num{6447.186}$^{+\num{11.9014901703}\%}_{- \num{11.7009208229 }\%}$ \\ (\num{7.51401371424}$\%)$}	
				        & \shortstack{\\ \num{6831.59242}$^{+\num{12.1751881057}\%}_{- \num{12.5244045123 }\%}$ \\ (\num{7.50565313356}$\%)$}	\\
\hline
+$m_{\ell\ell}$ Requirements 	&  \shortstack{\\ \num{589.98368}$^{+\num{4.15463745574}\%}_{- \num{4.84261961588 }\%}$ \\ (\num{69.152827342}$\%)$} 
						& \shortstack{\\ \num{587.90321}$^{+\num{4.52566297653}\%}_{- \num{5.16211380834 }\%}$ \\ (\num{69.09226909}$\%)$} 
						& \shortstack{\\ \num{578.16441317}$^{+\num{5.29296008885}\%}_{- \num{5.45827640306 }\%}$ \\ (\num{68.5878346753}$\%)$} 
						& \shortstack{\\ \num{39.659833}$^{+\num{7.00582443333}\%}_{- \num{7.56238349305 }\%}$ \\ (\num{28.3356954581}$\%)$}
						& \shortstack{\\ \num{39.617371}$^{+\num{6.88171173603}\%}_{- \num{7.35446923045 }\%}$ \\ (\num{28.3212686834}$\%)$}
						& \shortstack{\\ \num{40.058232336}$^{+\num{6.33593309563}\%}_{- \num{6.79268040677 }\%}$ \\ (\num{28.7739012425}$\%)$}
						&  \shortstack{\\ \num{4992.099}$^{+\num{11.9004898713}\%}_{- \num{11.7118592732 }\%}$ \\ (\num{77.453817333}$\%)$}
						& \shortstack{\\ \num{5000.659}$^{+\num{11.9029393831}\%}_{- \num{11.700934056 }\%}$ \\ (\num{77.5634478885}$\%)$}
						& \shortstack{\\ \num{5302.376586}$^{+\num{12.3594162614}\%}_{- \num{13.3792100516 }\%}$ \\ (\num{77.6155253378}$\%)$} \\
\hline
+Minimum $M_{T2}$             	&  \shortstack{\\ \num{2.9163611}$^{+\num{6.59339733889}\%}_{- \num{7.48118604276 }\%}$ \\ (\num{0.494312189545}$\%)$}
						& \shortstack{\\ \num{2.970984}$^{+\num{4.58976277801}\%}_{- \num{8.78621230925 }\%}$ \\ (\num{0.50535245405}$\%)$}
						& \shortstack{\\ \num{2.40445571}$^{+\num{8.99960371706}\%}_{- \num{6.22101978059 }\%}$ \\ (\num{0.41587752135}$\%)$}
						& \shortstack{\\ \num{0.74026389}$^{+\num{9.56728487965}\%}_{- \num{10.1762346862 }\%}$ \\ (\num{1.86653328721}$\%)$}
						& \shortstack{\\ \num{0.75861749}$^{+\num{7.19844720826}\%}_{- \num{7.70144743995 }\%}$ \\ (\num{1.9148607608}$\%)$}
						& \shortstack{\\ \num{0.5565072468}$^{+\num{6.97698125832}\%}_{- \num{8.20950326932 }\%}$ \\ (\num{1.38924573316}$\%)$}
						&  \shortstack{\\ \num{8.823189}$^{+\num{13.0867797561}\%}_{- \num{14.7501711858 }\%}$ \\ (\num{0.176743112459}$\%)$}
						& \shortstack{\\ \num{9.234719}$^{+\num{13.1559076653}\%}_{- \num{12.4020273581 }\%}$ \\ (\num{0.184670079168}$\%)$}
						& \shortstack{\\ \num{8.585987396}$^{+\num{12.4304452634}\%}_{- \num{25.2035076755 }\%}$ \\ (\num{0.161927165254}$\%)$}\\
\hline
\multicolumn{10}{c}{Benchmark Static Jet Veto Analysis Requirements} \\
\hline
+\mbox{$p_T^{\ell_1} \!>\! 50\GeV,$}
$p_T^{\ell_2} > 20\GeV$       	& \shortstack{\\ \num{2.7477437}$^{+\num{9.04107814584}\%}_{- \num{8.87013166844 }\%}$ \\ (\num{94.2182410423}$\%)$}
						& \shortstack{\\ \num{2.8047414}$^{+\num{4.60441727517}\%}_{- \num{11.9974744735 }\%}$ \\ (\num{94.4044764189}$\%)$}
						& \shortstack{\\ \num{2.2986718035}$^{+\num{11.657635804}\%}_{- \num{7.15371324021 }\%}$ \\ (\num{95.6004756243}$\%)$}
						& \shortstack{\\ \num{0.70059031}$^{+\num{11.1849108382}\%}_{- \num{11.9295136616 }\%}$ \\ (\num{94.6406210869}$\%)$}
						& \shortstack{\\ \num{0.71412389}$^{+\num{7.41174482747}\%}_{- \num{7.9820521686 }\%}$ \\ (\num{94.1348973607}$\%)$}
						& \shortstack{\\ \num{0.5275914249}$^{+\num{7.81490601507}\%}_{- \num{10.2622112468 }\%}$ \\ (\num{94.8040510788}$\%)$}
						&  \shortstack{\\ \num{7.670907}$^{+\num{13.1386593917}\%}_{- \num{19.9771899867 }\%}$ \\ (\num{86.9402985075}$\%)$}
						& \shortstack{\\ \num{8.296435}$^{+\num{14.6373669361}\%}_{- \num{12.6801369704 }\%}$ \\ (\num{89.8395721925}$\%)$}
						& \shortstack{\\ \num{7.79873434}$^{+\num{12.4867806937}\%}_{- \num{32.8379375823 }\%}$ \\ (\num{90.8309455587}$\%)$}\\
\hline                              
+Static Jet Veto              	&  \shortstack{\\ \num{1.4748052}$^{+\num{13.47297633}\%}_{- \num{11.2056948852 }\%}$ \\ (\num{53.6732929991}$\%)$}
					& \shortstack{\\ \num{1.479555}$^{+\num{5.61173843962}\%}_{- \num{17.3626071467 }\%}$ \\ (\num{52.7519051651}$\%)$}
					& \shortstack{\\ \num{1.6125009129}$^{+\num{12.7499258412}\%}_{- \num{7.77286511754 }\%}$ \\ (\num{70.1492537313}$\%)$}
					& \shortstack{\\ \num{0.27363621}$^{+\num{11.1860888881}\%}_{- \num{12.3958568804 }\%}$ \\ (\num{39.0579518391}$\%)$}
					& \shortstack{\\ \num{0.25899038}$^{+\num{9.26532984708}\%}_{- \num{8.01196785879 }\%}$ \\ (\num{36.266874351}$\%)$}
					& \shortstack{\\ \num{0.31415332036}$^{+\num{7.84065391789}\%}_{- \num{12.8558004909 }\%}$ \\ (\num{59.5448211797}$\%)$}
					&  \shortstack{\\ \num{0.03292236}$^{+\num{244.79191656}\%}_{- \num{20.434011745 }\%}$ \\ (\num{0.429184549356}$\%)$}
					& \shortstack{\\ \num{0.04938354}$^{+\num{65.6706039873}\%}_{- \num{66.4554120812 }\%}$ \\ (\num{0.595238095238}$\%)$}
					& \shortstack{\\ \num{0.1476100939}$^{+\num{68.3160353671}\%}_{- \num{57.6817750446 }\%}$ \\ (\num{1.8927444795}$\%)$} \\

\hline
\multicolumn{10}{c}{Dynamic Jet Veto Analysis Requirements} \\
\hline

$\pTVeto = p_T^{\ell_1}$    	& \shortstack{\\ \num{2.581502}$^{+\num{6.65091716741}\%}_{- \num{14.1575978346 }\%}$ \\ (\num{86.4757358791}$\%)$}
						& \shortstack{\\ \num{2.4888812}$^{+\num{7.02309159575}\%}_{- \num{8.84348755467 }\%}$ \\ (\num{85.4812398042}$\%)$}
						&  \shortstack{\\ \num{2.4130591265}$^{+\num{10.2082694945}\%}_{- \num{7.40559882998 }\%}$ \\ $(>99\%)$}
						& \shortstack{\\ \num{0.58342363}$^{+\num{10.5823430932}\%}_{- \num{9.8566451001 }\%}$ \\ (\num{77.4169741697}$\%)$}
						& \shortstack{\\ \num{0.58194045}$^{+\num{10.8247901995}\%}_{- \num{10.1459898144 }\%}$ \\ (\num{77.1821981805}$\%)$}
						& \shortstack{\\ \num{0.55405683892}$^{+\num{7.29454017162}\%}_{- \num{8.58258111114 }\%}$ \\ (\num{98.7336244541}$\%)$}
						& \shortstack{\\ \num{3.786071}$^{+\num{12.1240786918}\%}_{- \num{12.9147179609 }\%}$ \\ (\num{42.5925925926}$\%)$}
						& \shortstack{\\ \num{3.851916}$^{+\num{12.3712247843}\%}_{- \num{13.4968587571 }\%}$ \\ (\num{41.6370106762}$\%)$}
						& \shortstack{\\ \num{4.526709449}$^{+\num{31.868970568}\%}_{- \num{32.2454879282 }\%}$ \\ (\num{57.1428571429}$\%)$} \\
\hline
$\pTVeto = p_T^{\ell_2}$    	& \shortstack{\\ \num{1.9521567}$^{+\num{4.36101844488}\%}_{- \num{7.98889701793 }\%}$ \\ (\num{67.1568627451}$\%)$}
						&  \shortstack{\\ \num{1.9474073}$^{+\num{7.62034185206}\%}_{- \num{10.5303571269 }\%}$ \\ (\num{67.7685950413}$\%)$}
						& \shortstack{\\ \num{2.2043229213}$^{+\num{11.2360287797}\%}_{- \num{8.6262088753 }\%}$ \\ (\num{90.7058823529}$\%)$}
						& \shortstack{\\ \num{0.40581959}$^{+\num{14.7252977439}\%}_{- \num{12.9562023982 }\%}$ \\ (\num{55.0138225685}$\%)$}
						& \shortstack{\\ \num{0.4265833}$^{+\num{9.10158292579}\%}_{- \num{8.87732691291 }\%}$ \\ (\num{56.7867719645}$\%)$}
						& \shortstack{\\ \num{0.5043116946}$^{+\num{7.07677405428}\%}_{- \num{7.51598478043 }\%}$ \\ (\num{87.4256584537}$\%)$}
						& \shortstack{\\ \num{0.8395198}$^{+\num{24.9161678452}\%}_{- \num{34.9875618895 }\%}$ \\ (\num{8.99470899471}$\%)$}
						& \shortstack{\\ \num{0.9547485}$^{+\num{22.2832559537}\%}_{- \num{11.7835629316 }\%}$ \\ (\num{10.3202846975}$\%)$}
						& \shortstack{\\ \num{0.9840669036}$^{+\num{57.4015022484}\%}_{- \num{16.8611721484 }\%}$ \\ (\num{12.3456790123}$\%)$} \\
\hline
$\pTVeto = S_T$             		& \shortstack{\\ \num{2.7738678}$^{+\num{6.17103524041}\%}_{- \num{11.8080082457 }\%}$ \\ (\num{94.2695722357}$\%)$}
						&  \shortstack{\\ \num{2.6978712}$^{+\num{8.15053930177}\%}_{- \num{7.71421609857 }\%}$ \\ (\num{94.1956882255}$\%)$}
						&  \shortstack{\\ \num{2.4301874077}$^{+\num{8.85144998639}\%}_{- \num{6.7051725528 }\%}$ \\ $(>99\%)$}
						& \shortstack{\\ \num{0.67500645}$^{+\num{8.8592164414}\%}_{- \num{9.92392961355 }\%}$ \\ (\num{89.701897019}$\%)$}
						& \shortstack{\\ \num{0.66740542}$^{+\num{9.41097786639}\%}_{- \num{9.49926067228 }\%}$ \\ (\num{89.6860986547}$\%)$}
						& \shortstack{\\ \num{0.56434881368}$^{+\num{6.86376240778}\%}_{- \num{8.97045275909 }\%}$ \\ $(>99\%)$}
						& \shortstack{\\ \num{5.794335}$^{+\num{18.7249520416}\%}_{- \num{18.0692579044 }\%}$ \\ (\num{61.9718309859}$\%)$}
						& \shortstack{\\ \num{7.061846}$^{+\num{13.0374602765}\%}_{- \num{25.6094004521 }\%}$ \\ (\num{70.9090909091}$\%)$}
						& \shortstack{\\ \num{6.568649981}$^{+\num{23.3694762069}\%}_{- \num{26.0812741224 }\%}$ \\ (\num{89.0}$\%)$} \\
\hline
$\HTVeto = p_T^{\ell_1}$          & \shortstack{\\ \num{2.1872713}$^{+\num{7.43756900897}\%}_{- \num{15.8475975952 }\%}$ \\ (\num{73.2114467409}$\%)$}
						& \shortstack{\\ \num{1.8809104}$^{+\num{10.4186328244}\%}_{- \num{7.29052302613 }\%}$ \\ (\num{67.1755725191}$\%)$}
						& \shortstack{\\ \num{2.1585785293}$^{+\num{9.20537773464}\%}_{- \num{5.831921021 }\%}$ \\ (\num{86.8814729574}$\%)$}
						& \shortstack{\\ \num{0.45568959}$^{+\num{9.41563455622}\%}_{- \num{8.67761130899 }\%}$ \\ (\num{60.4971695791}$\%)$}
						& \shortstack{\\ \num{0.41045434}$^{+\num{16.1735389468}\%}_{- \num{10.3751019385 }\%}$ \\ (\num{55.4470323065}$\%)$}
						& \shortstack{\\ \num{0.46216324746}$^{+\num{7.27869374709}\%}_{- \num{12.858840443 }\%}$ \\ (\num{83.896797153}$\%)$}
						& \shortstack{\\ \num{0.3950683}$^{+\num{23.3322022617}\%}_{- \num{12.6186312725 }\%}$ \\ (\num{4.24028268551}$\%)$}
						& \shortstack{\\ \num{0.4938354}$^{+\num{24.6624427792}\%}_{- \num{50.1074950072 }\%}$ \\ (\num{4.89396411093}$\%)$}
						& \shortstack{\\ \num{0.54123695747}$^{+\num{21.8013149608}\%}_{- \num{24.707533833 }\%}$ \\ (\num{7.07395498392}$\%)$} \\
\hline
$\HTVeto = p_T^{\ell_2}$          & \shortstack{\\ \num{1.4795543}$^{+\num{4.8503380487}\%}_{- \num{10.5175879198 }\%}$ \\ (\num{51.1914543961}$\%)$}
						& \shortstack{\\ \num{1.2230672}$^{+\num{5.28692246299}\%}_{- \num{5.44025119564 }\%}$ \\ (\num{42.9524603837}$\%)$}
						& \shortstack{\\ \num{1.45525295485}$^{+\num{7.69973804463}\%}_{- \num{7.30106136916 }\%}$ \\ (\num{59.8823529412}$\%)$}
						& \shortstack{\\ \num{0.30923118}$^{+\num{8.29124794432}\%}_{- \num{8.61629479487 }\%}$ \\ (\num{40.9626719057}$\%)$}
						& \shortstack{\\ \num{0.25398487}$^{+\num{8.94171912104}\%}_{- \num{8.03027147468 }\%}$ \\ (\num{33.5455435847}$\%)$}
						& \shortstack{\\ \num{0.3308167632}$^{+\num{8.61296842247}\%}_{- \num{8.90133090507 }\%}$ \\ (\num{57.9896907216}$\%)$}
						& \shortstack{\\ \num{0.04938354}$^{+\num{69.1418000844}\%}_{- \num{12.976704197 }\%}$ \\ (\num{0.528169014085}$\%)$}
						& \shortstack{\\ \num{0.01646118}$^{+\num{195.635851353}\%}_{- \num{213.217123948 }\%}$ \\ (\num{0.170648464164}$\%)$}
						&  \shortstack{\\ \num{0.12300842131}$^{+\num{61.8910587253}\%}_{- \num{81.3272426063 }\%}$ \\ (\num{1.53374233129}$\%)$} \\
\hline
$\HTVeto = S_T$                   	& \shortstack{\\ \num{2.4817573}$^{+\num{4.63237688064}\%}_{- \num{13.1332174136 }\%}$ \\ (\num{83.733974359}$\%)$}
						& \shortstack{\\ \num{2.3796363}$^{+\num{6.09358253431}\%}_{- \num{8.3375836144 }\%}$ \\ (\num{82.1985233798}$\%)$}
						&  \shortstack{\\ \num{2.3301204765}$^{+\num{9.57166846048}\%}_{- \num{9.6539282394 }\%}$ \\ (\num{96.7933491686}$\%)$}
						& \shortstack{\\ \num{0.55098032}$^{+\num{12.9114561049}\%}_{- \num{10.3110221098 }\%}$ \\ (\num{74.392991239}$\%)$}
						& \shortstack{\\ \num{0.5309581}$^{+\num{10.601120095}\%}_{- \num{10.3753703551 }\%}$ \\ (\num{71.3325031133}$\%)$}
						& \shortstack{\\ \num{0.53690331702}$^{+\num{6.98133935801}\%}_{- \num{8.47949667198 }\%}$ \\ (\num{95.1367781155}$\%)$}
						& \shortstack{\\ \num{1.4321224}$^{+\num{23.0110707206}\%}_{- \num{24.0129252557 }\%}$ \\ (\num{15.5080213904}$\%)$}
						& \shortstack{\\ \num{1.1028985}$^{+\num{44.1133008172}\%}_{- \num{14.8197794959 }\%}$ \\ (\num{11.2984822934}$\%)$}
						& \shortstack{\\ \num{1.8943298545}$^{+\num{30.8627886418}\%}_{- \num{20.4656881304 }\%}$ \\ (\num{23.1231231231}$\%)$} \\
\hline
\hline
\end{tabular}
}
\caption{
The cross section [fb] with uncertainties [$^{+\%}_{-\%}$] and cut efficiency [(\%)] of the selection cuts in table~\ref{tb:SelectionCuts14TeV}
for the dominant SM backgrounds, when modeled at the inclusive NLO+PS level without MPI (Incl.), with MPI (Incl.+MPI), and FxFx-merging with MPI (FxFx+MPI).
Uncertainties are obtained by adding the renormalization and factorization scale envelope with the shower scale envelope and statistical uncertainty in quadrature.
At the generator-level, statistical confidence corresponds to 5-10 M events for each sample and shower variation.
\label{tab:CMSCutflows}
}
\end{table*}

\begin{figure*}
\centering
\subfigure[]{\includegraphics[width=.48\textwidth]{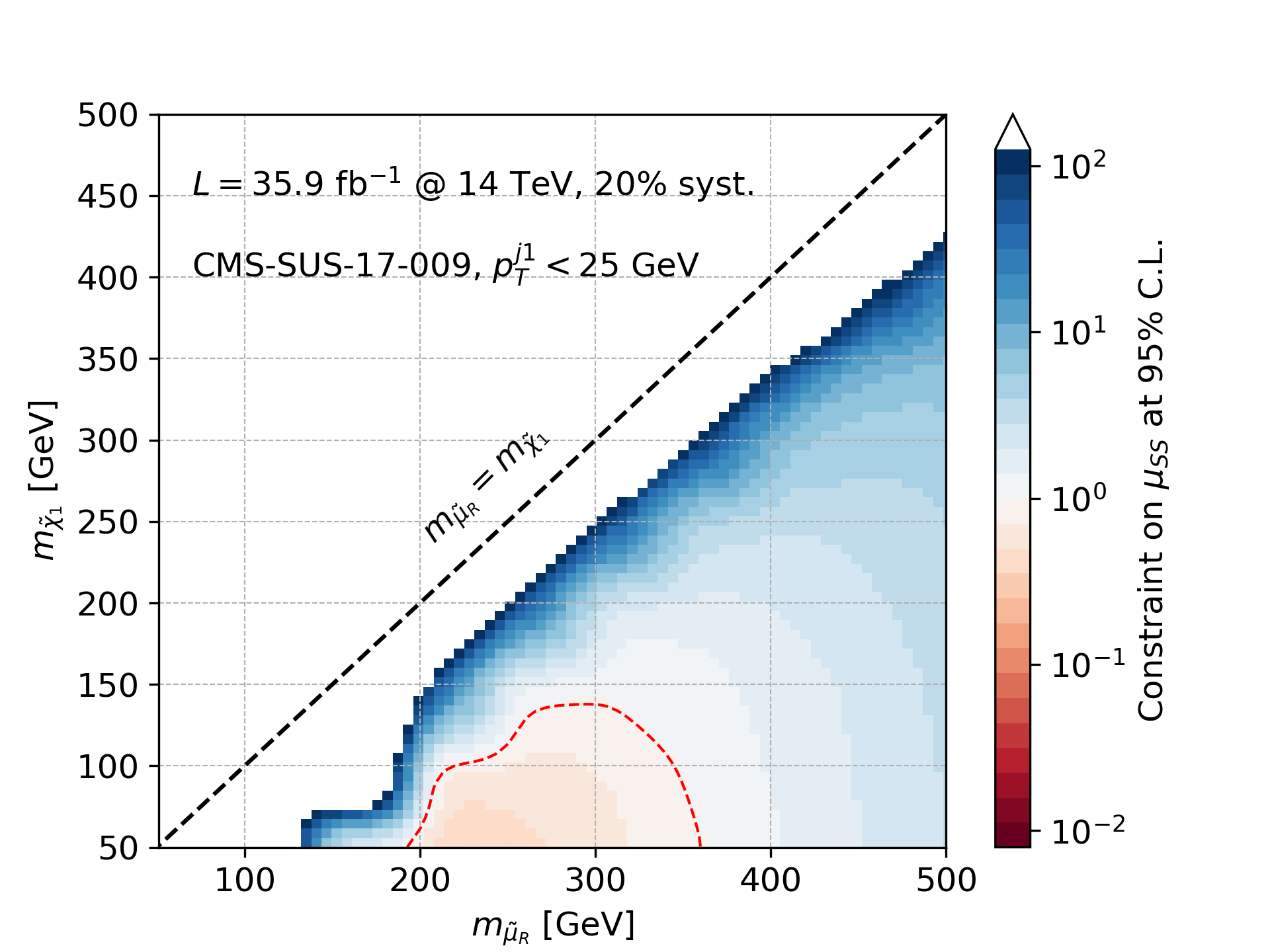}	}
\subfigure[]{\includegraphics[width=.48\textwidth]{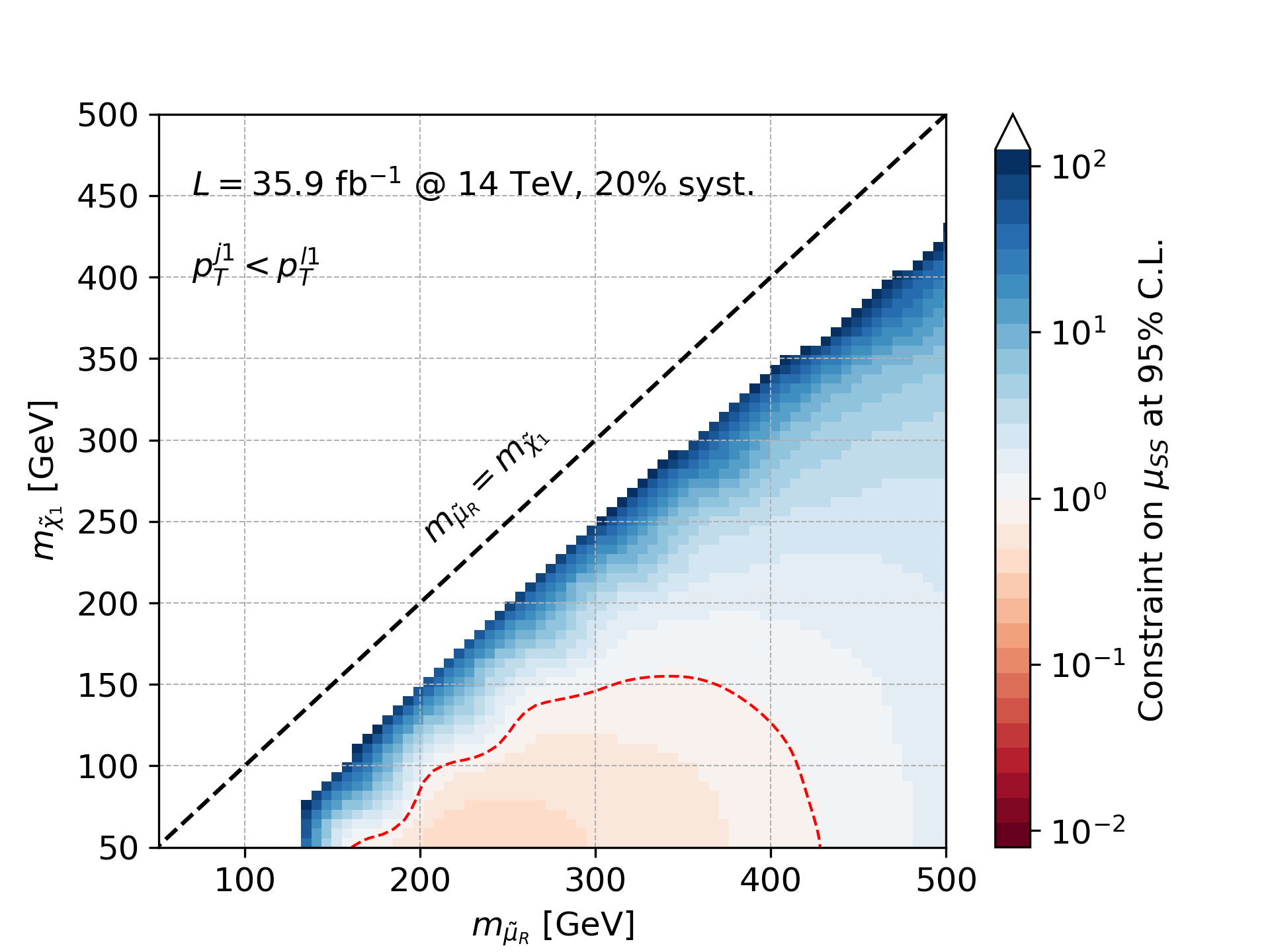}	}
\caption{Exclusion contours on the signal strength $\mu_{SS}$ for smuon pair production in the $(m_{\tilde{\chi}_1},m_{\tilde{\mu}_R})$ plane with $\mathcal{L}=35.9\invfb$,
for (a) the static jet veto analysis based on ref.~\cite{Sirunyan:2018nwe}, with $\pTVeto=25\GeV$,
and
(b) the dynamic jet veto analysis with $\pTVeto = p_{T}^{\ell_1}$.
}
\label{fig:sleptonVeto_95CLSensitivity_LHC14_36invfb}
\end{figure*}

\subsubsection*{Dynamic Jet Veto Analysis}\label{sec:dynamicAnalysis}
The goal of this study is to see to what extent, if at all, generalizations of dynamic jet vetoes  can improve searches for multilepton final states over traditional, static, central jet vetoes.
To do this, we propose a class of analyses that simplifies the static veto analysis of the preceding subsection.
We execute this by removing the stringent high-$p_T$ selection cuts on charged leptons given in eq.~(\ref{cut:lepPT}) and by setting the central jet veto threshold on an event-by-event basis.
As can be observed from table~\ref{tab:CMSCutflows}, these lepton-$p_T$ cuts only reduce backgrounds by 5-10\% but at the cost of hindering signal acceptance, particularly in the mass-degenerate limit.
More precisely, 	events are vetoed either (i) if there exists an analysis-quality jet with $p_T^j > \pTVeto$ or (ii) if the event possess $H_T > \HTVeto$.
In no case do we consider simultaneously a veto on $p_T^j$ and on $H_T$.
The veto thresholds are set dynamically according to the following permutations:
\begin{eqnarray}
{\rm (a)}~\pTVeto = p_T^{\ell_1},	\quad		&{\rm (b)}~\HTVeto = p_T^{\ell_1}, &		\quad	{\rm (c)}~\pTVeto = S_T,			\nonumber\\
{\rm (d)}~\HTVeto = S_T,			\quad		&{\rm (e)}~\pTVeto = p_T^{\ell_2}, &		\quad	{\rm (f)}~\HTVeto = p_T^{\ell_2}.		\nonumber
\end{eqnarray}
In principle, one can introduce a scaling factor $r$, {\it e.g.}, $\HTVeto = r \times S_T$, with $r = 0.75$, and improve the signal-to-background ratio $S/B$ according to fig.~\ref{fig:DynamicRatios}.
However, this is beyond the proof-of-concept scope of our study.
Needless to say, investigations into optimizing a ``smart jet veto" are encouraged.

\section{Results and Outlook}\label{sec:results}

To quantify the impact of dynamic jet vetoes on searches for smuon pairs,
we use the $CL_S$ technique \cite{Read:2002hq} to first determine the 95\% CL reach
in terms of the event rate $N_{95} = \sigma_{95} \times \mathcal{L}$, for a luminosity $\mathcal{L}$.
We take into account the Monte Carlo uncertainties for both the signal and the background, and use a flat systematic uncertainty of 20\% on the background prediction derived from our FxFx + MPI samples.
We use the combined likelihood ratio of the four signal regions as our test statistic.
Sensitivity is then expressed in terms of the signal strength $(\mu_{SS})$,
\begin{equation}
\mu_{SS} = \sigma_{95}/\sigma_p,
\end{equation}
where $\sigma_p$ is the predicted cross section in our simplified model.
A signal strength of $\mu_{SS}<1$ means that the signal hypothesis is excluded with at least 95\% confidence.

As a check, we show in fig.~\ref{fig:sleptonVeto_95CLSensitivity_LHC14_36invfb}, $\mu_{SS}$
for (a)  the static jet veto analysis based on ref.~\cite{Sirunyan:2018nwe}, where $\pTVeto=25\GeV$,
and (b) the dynamic jet veto $\pTVeto = p_{T}^{\ell_1}$, assuming $\mathcal{L}=35.9\invfb$ at $\sqrt{s}=14\TeV$.
To derive these we only consider the Monte Carlo uncertainty and a flat 20\% additional systematic uncertainty
(which is intended to approximate all additional theory and experimental systematic uncertainties) in the limit setting, to keep the comparison as clear as possible.
We find that the constraints derived using the reference analysis are stronger than those reported in ref.~\cite{Sirunyan:2018nwe}.
This is attributed to three reasons.
First, in comparison with the 13~TeV results explored with data, we take $\sqrt{s}=14\TeV$.
Second, we use a highly simplified treatment of systematic uncertainties, and finally,
we recover a slightly smaller background prediction for the lowest $\slashed{E}_T$ signal region compared with the data-driven prediction of ref.~\cite{Sirunyan:2018nwe} (see sec.~\ref{sec:jetVetoAna}).

With the dynamic jet veto analysis, we
observe an improvement in sensitivity over the static veto analysis,
with $m_{\tilde{\mu}_R}\lesssim425\GeV$ being accessible for $m_{\tilde{\mu}_R}\gg m_{\tilde{\chi}_1}$, to be confronted to $m_{\tilde{\mu}_R}\lesssim 360\GeV$ in the static case.
For larger luminosities we find that the improvement is comparable.
However, as part of the improvement comes from higher signal acceptance rather than large improvements in $S/B$, the relative improvement diminishes somewhat.
We stress that while this improvement appears limited,
it has been obtained by relaxing several selection cuts of the somewhat sophisticated analysis of ref.~\cite{Sirunyan:2018nwe},
and na\"ively applying a dynamic jet veto that has not been optimized according to fig.~\ref{fig:DynamicRatios}.
This ``out-of-the-box'' improvement even for relatively light smuon masses is encouraging.

\begin{figure*}
\centering
	\subfigure[]{\includegraphics[width=.48\textwidth]{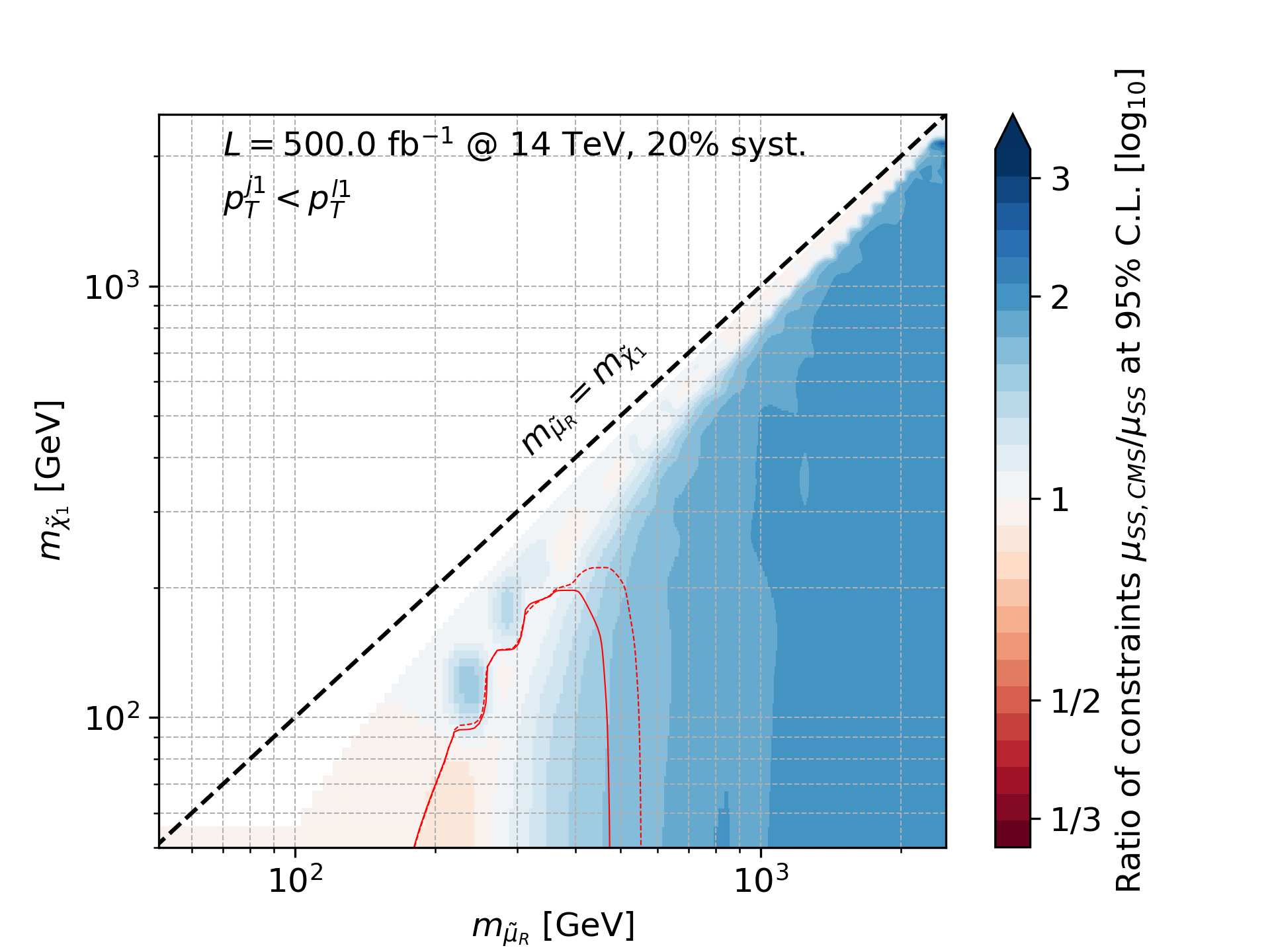}	}
	\subfigure[]{\includegraphics[width=.48\textwidth]{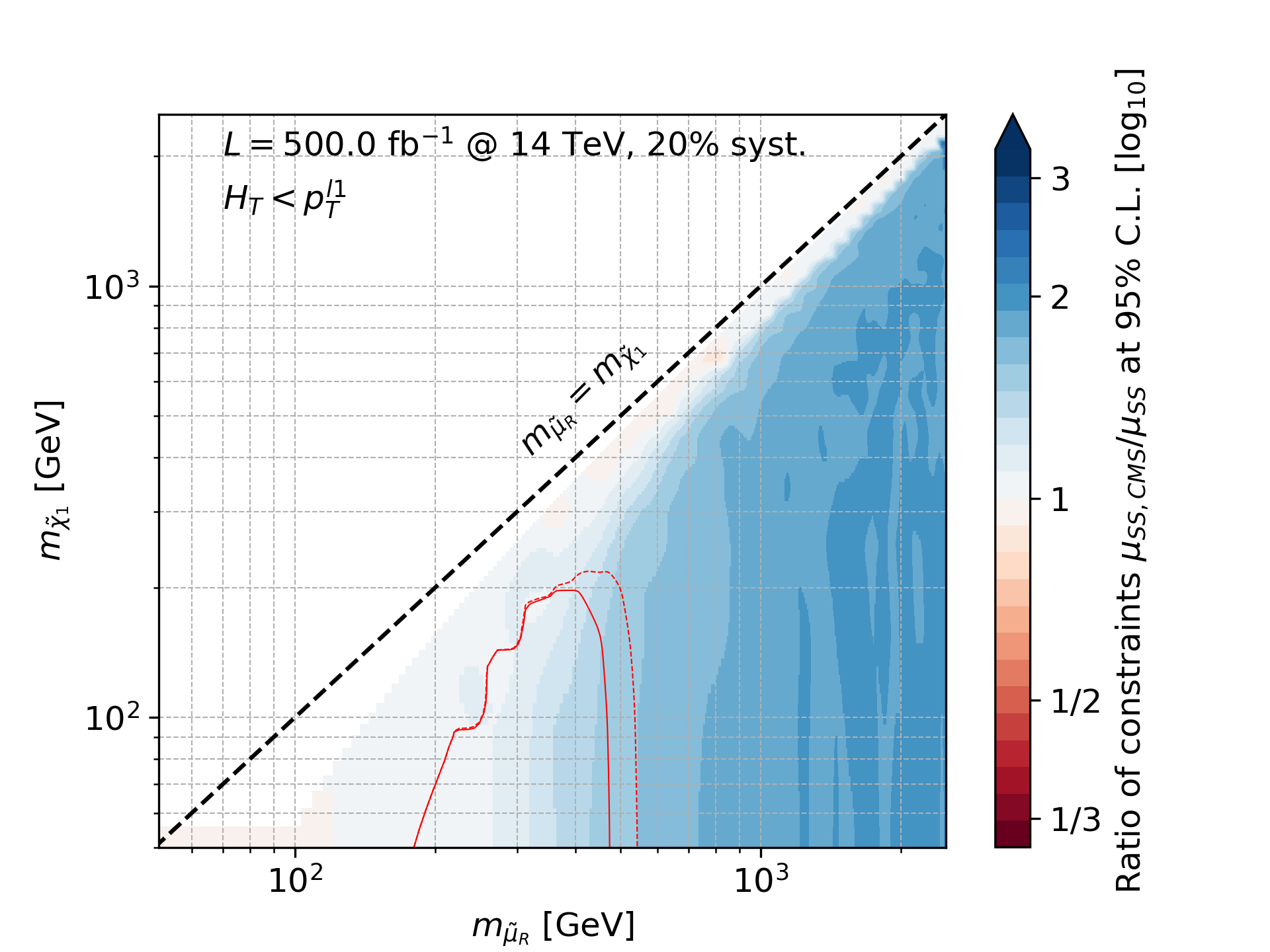}	}
	\\
	\subfigure[]{\includegraphics[width=.48\textwidth]{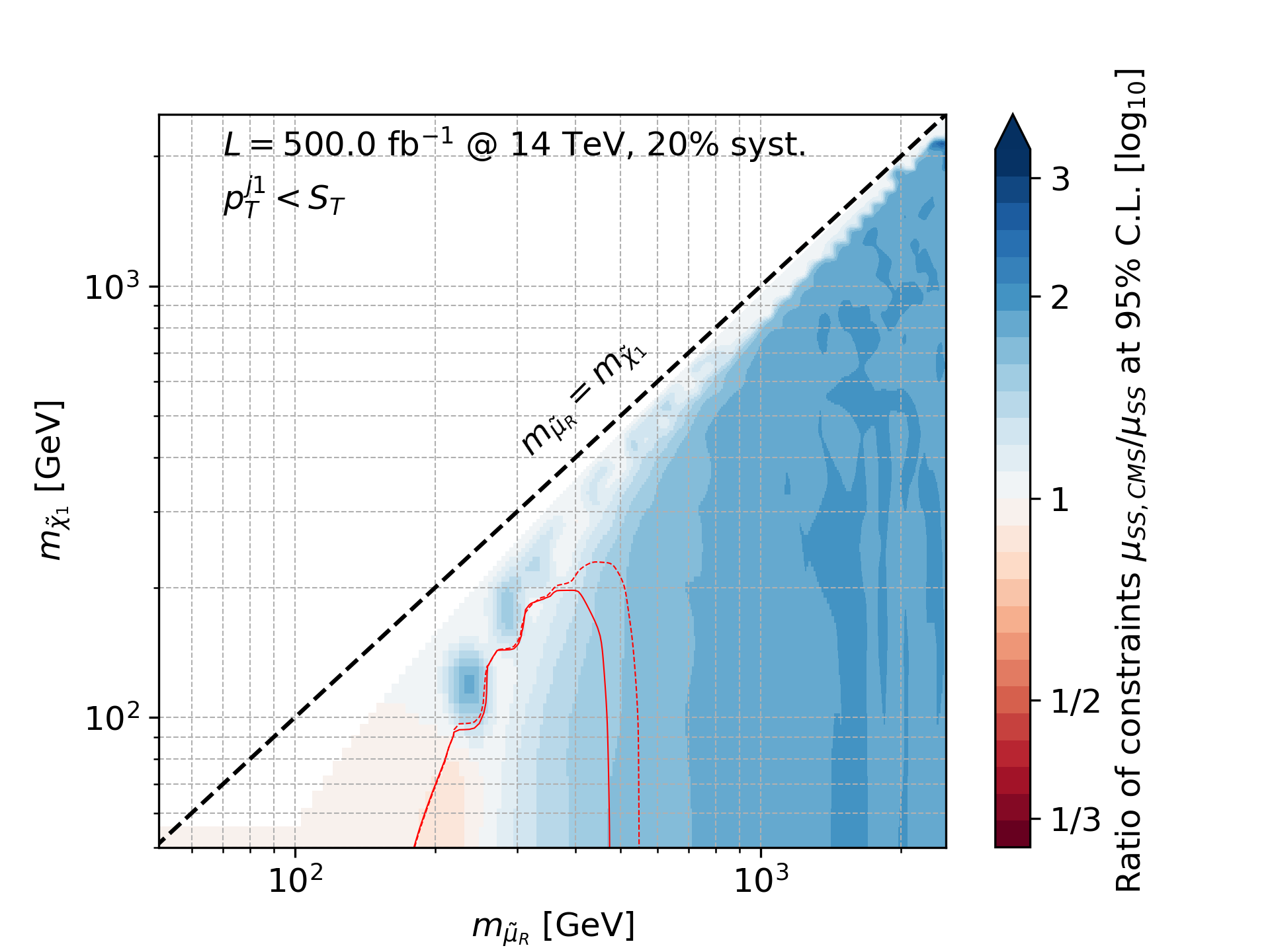}	}
	\subfigure[]{\includegraphics[width=.48\textwidth]{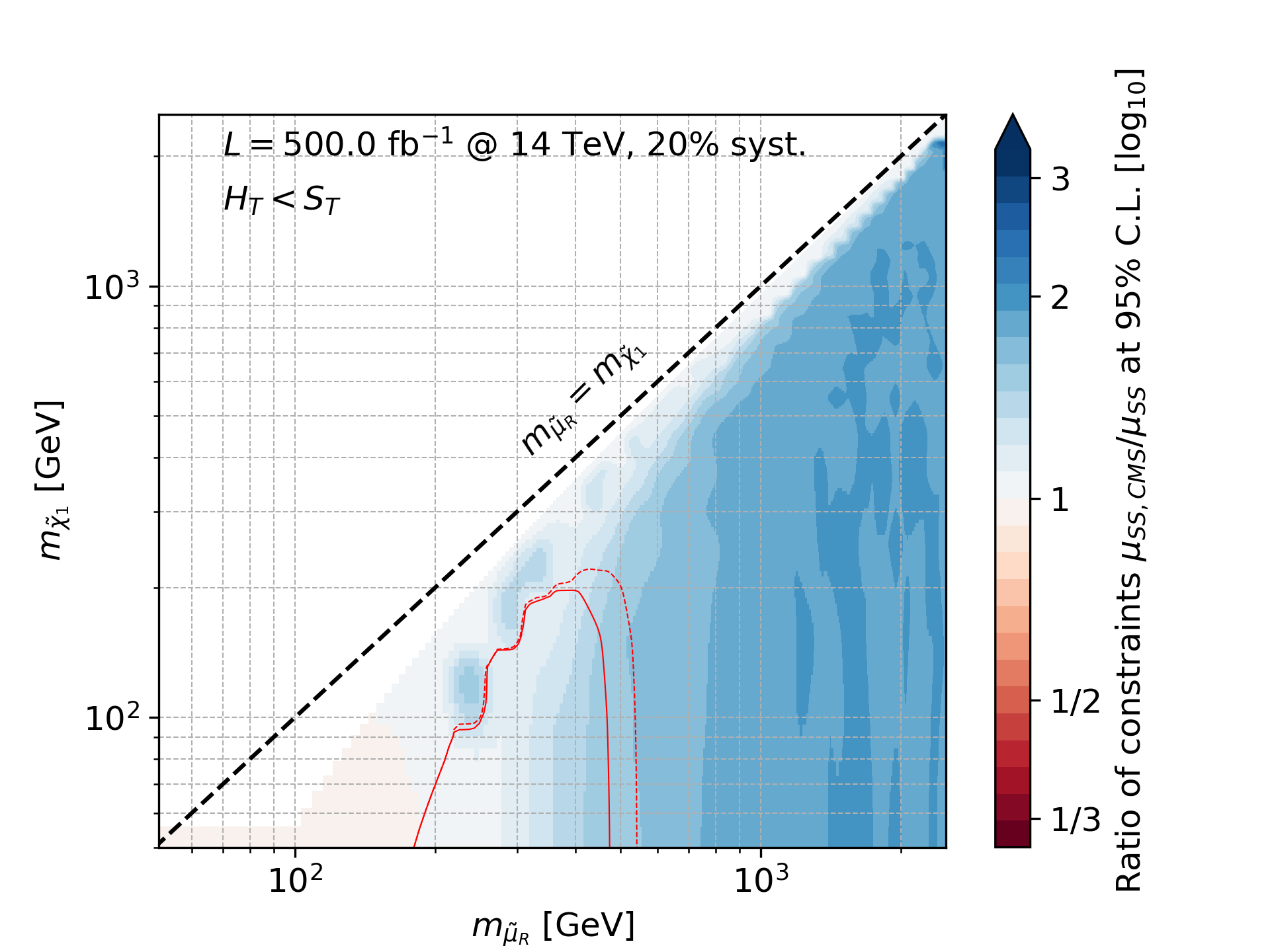}	}
	\\
	\subfigure[]{\includegraphics[width=.48\textwidth]{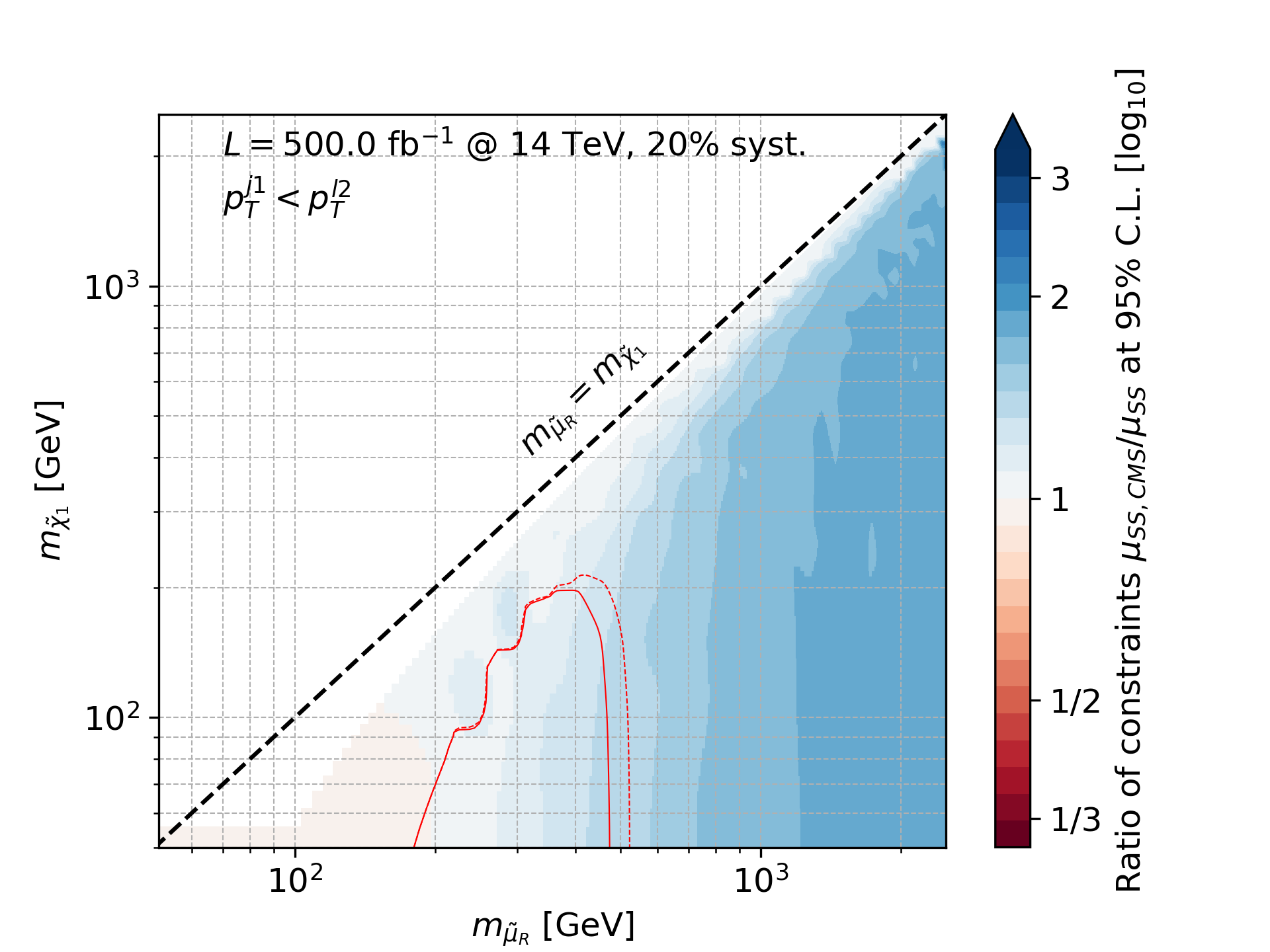}	}
  	\subfigure[]{\includegraphics[width=.48\textwidth]{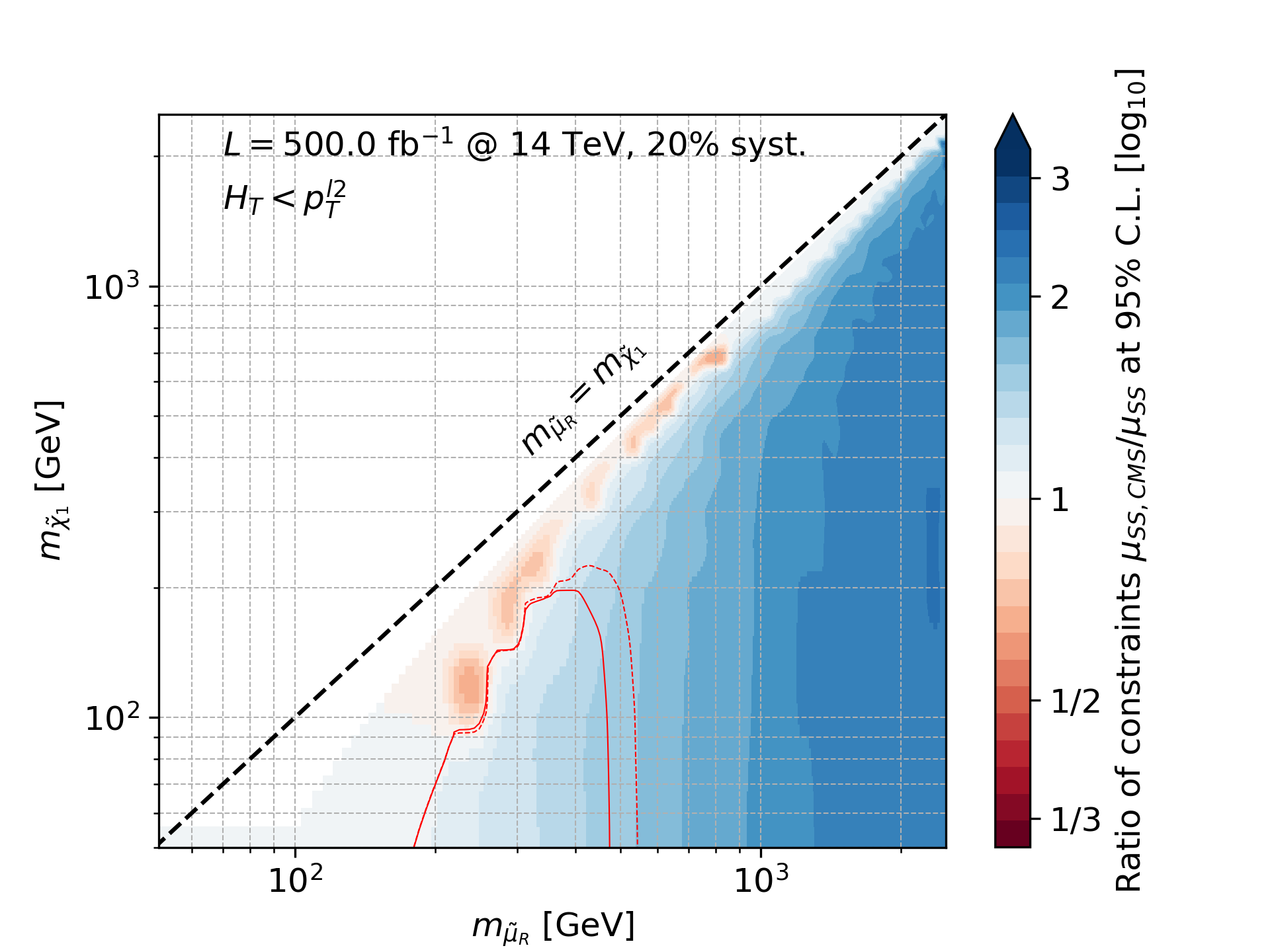}	}\\
\caption{The ratio of signal strengths $(\mu_{SS})$ for
(a) $\pTVeto = p_{T}^{\ell_1}$,
(b) $\HTVeto = p_{T}^{\ell_1}$,
(c) $\pTVeto = S_T$,
(d) $\HTVeto = S_T$,
(e) $\pTVeto = p_{T}^{\ell_2}$,
and
(f) $\HTVeto = p_{T}^{\ell_2}$,
 compared with the CMS reference analysis using $\mathcal{L}=500\invfb$. The solid red line shows the 95\% exclusion for $\mu_{SS} = 1$ for the benchmark CMS analysis, and the dashed red line the same exclusion for the dynamic analysis.
 }
\label{fig:sleptonVeto_95CLSensitivity_LHC14_ratio}
\end{figure*}

\begin{figure*}
\centering
\subfigure[]{\includegraphics[width=.45\textwidth]{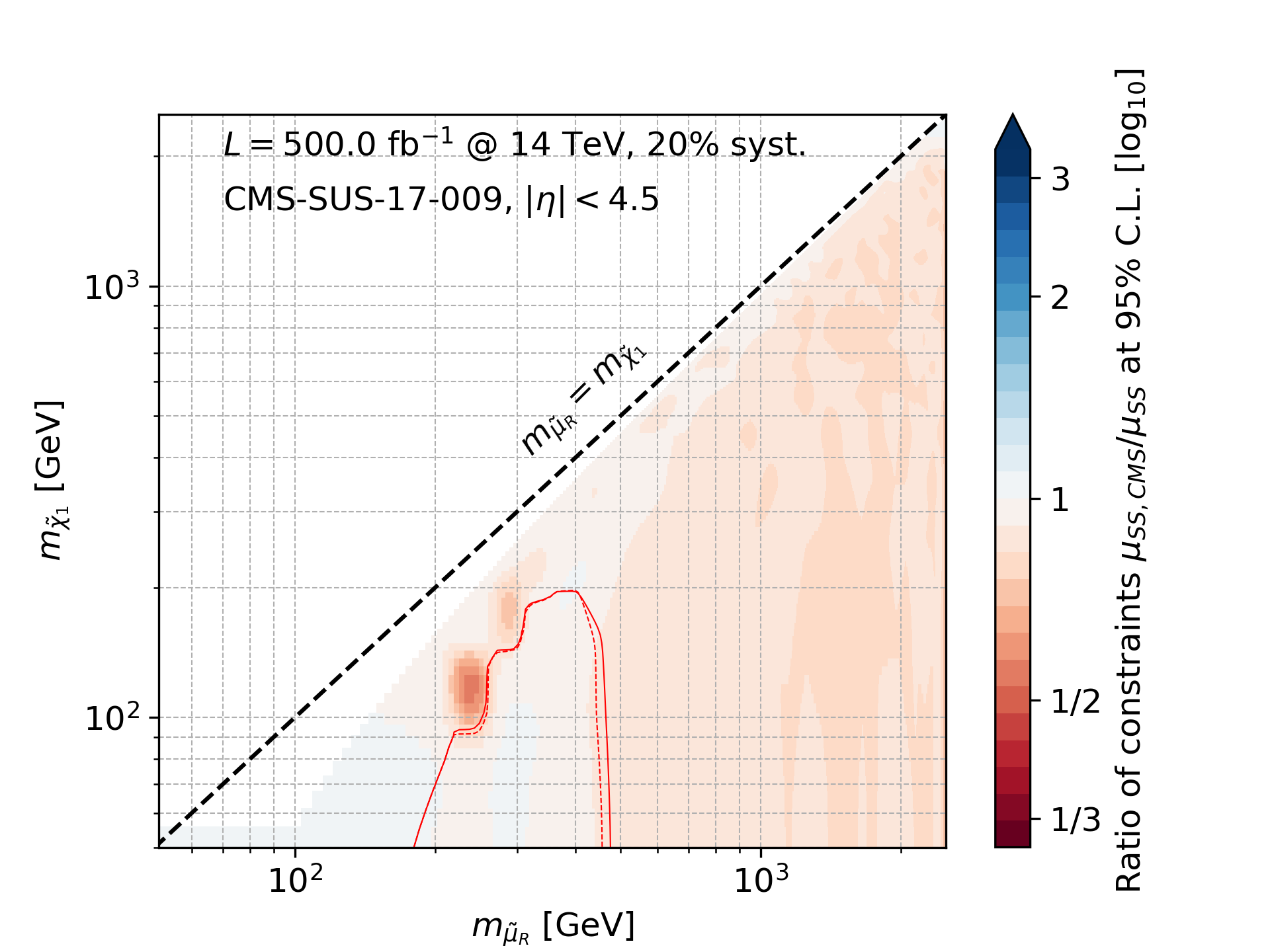}	}
\subfigure[]{\includegraphics[width=.45\textwidth]{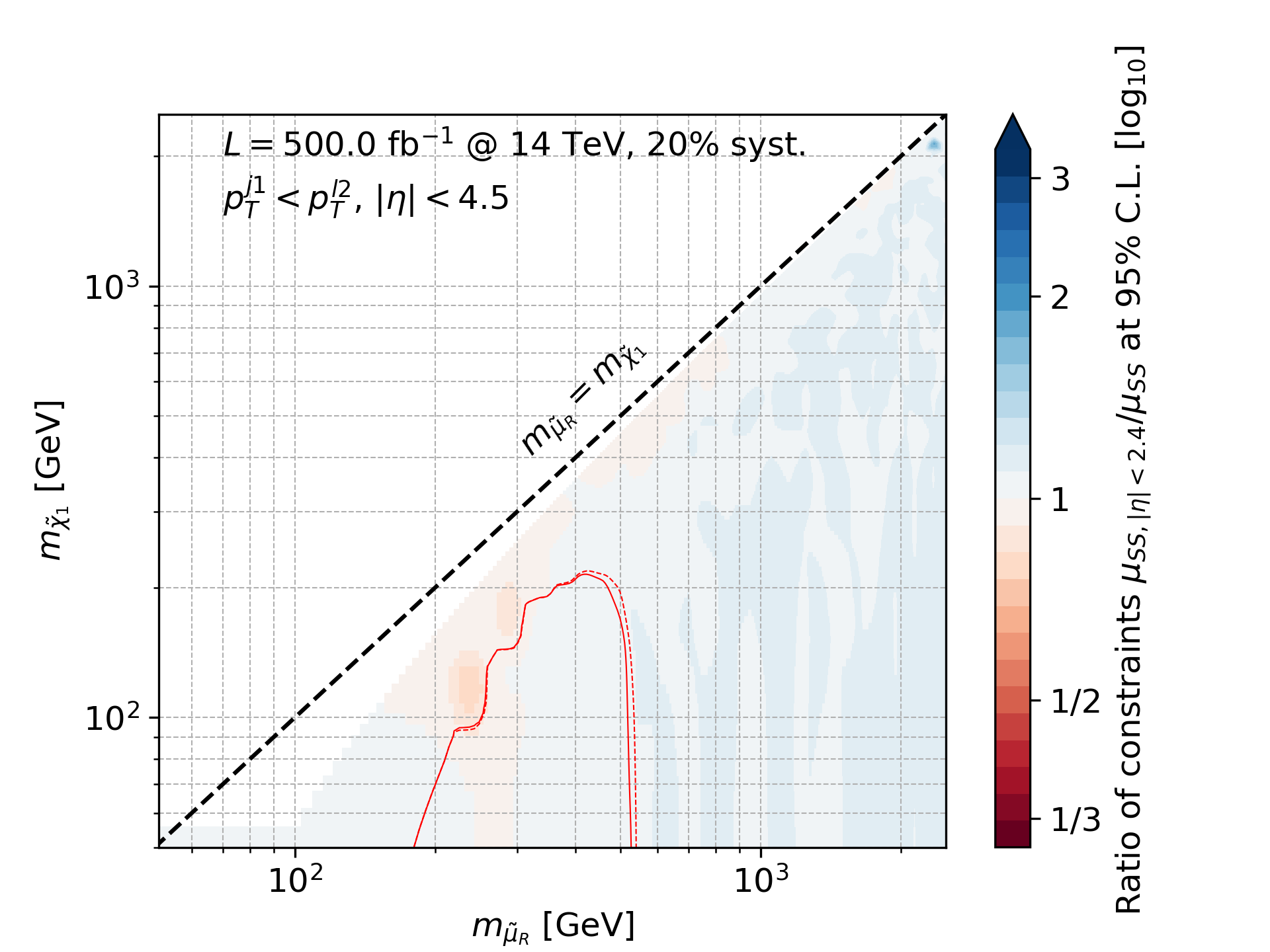}	}
\caption{The ratio of constraints at $\mathcal{L}=500\invfb$ for (a) the default CMS analysis, and (b) the dynamic $p_T^{j1} < p_T^{l2}$, both with a jet pseudorapidity cut of $|\eta| < 4.5$, compared with their respective default analyses employing $|\eta| < 2.4$. The solid red line shows the 95\% exclusion for $\mu_{SS} = 1$ for the standard $|\eta| < 2.4$ analysis, and the dashed red line the same exclusion for $|\eta| < 4.5$ analysis.}
\label{fig:sleptonVeto_Mu95Ratio_WideEta}
\end{figure*}

To present our main results, for a given jet veto scheme and luminosity we consider the ratio of signal strengths:
\begin{equation}
\mathcal{R}_{\rm Dy.~Veto} = \cfrac{\mu_{SS}^{\rm CMS}}{\mu_{SS}^{\rm Dy.~Veto}} = \cfrac{\sigma_{95}^{\rm CMS}/\sigma_p^{\rm CMS}}{\sigma_{95}^{\rm Dy.~Veto}/\sigma_p^{\rm Dy.~Veto}},
\end{equation}
where $\mu_{SS}^{\rm CMS}$ is the signal strength as determined using the reference static jet veto analysis
and $\mu_{SS}^{\rm Dy.~Veto}$ is the signal strength as determined with the dynamic jet veto analysis.
The double ratio has the simple interpretation that a value of  $\mathcal{R}>1$ implies that
the dynamic veto analysis is more sensitive than the static veto analysis for a given input.

In fig.~\ref{fig:sleptonVeto_95CLSensitivity_LHC14_ratio}, assuming $\mathcal{L}=500\invfb$, we present $\mathcal{R}$ for
\begin{eqnarray}
{\rm (a)} \pTVeto= p_{T}^{\ell_1}, \quad
{\rm (b)} \HTVeto= p_{T}^{\ell_1}, \quad
{\rm (c)} \pTVeto =  S_T,
\nonumber\\
{\rm (d)} \HTVeto = S_T, \quad
{\rm (e)} \pTVeto = p_{T}^{\ell_2}, \quad
{\rm (f)} \HTVeto = p_{T}^{\ell_2}.
\nonumber
\end{eqnarray}
In the large mass splitting regime where $m_{\tilde{\mu}_R}\gg m_{\tilde{\chi}_1}$,
we find that the veto scheme (f) $\HTVeto=p_T^{\ell_2}$
outperforms the static veto analysis for $m_{\tilde{\mu}_R}\gtrsim200\GeV$;
this finding extends to (b) $\HTVeto=p_T^{\ell_1}$, (d) $\HTVeto=S_T$ and (e) $\pTVeto=p_T^{\ell_2}$ for $m_{\tilde{\mu}_R}\gtrsim250\GeV$;
and we report that all dynamic jet veto schemes show improvement for $m_{\tilde{\mu}_R}\gtrsim300\GeV$.
Of the schemes considered, the choice $(c) \pTVeto = p_T^{\ell_2}$ arguably performs worst,
with limited improvement over the static analysis for much of the phenomenologically relevant parameter space.
For the compressed regime where $m_{\tilde{\mu}_R}\sim m_{\tilde{\chi}_1}$,
the $S_T$ schemes demonstrate some improvement,
while (f) $\HTVeto=p_T^{\ell_2}$ is considerably weaker than the static analysis.

For much of the parameter space of interest, we see that the improvement is in excess of $25\%$ to $50\%$.
The relative improvement grows with increasing $m_{\tilde{\mu}_R}$ which allows for improvement in excess of 100\%
since the static veto reduces the signal efficiency for heavier mass scales (due to harder initial-state radiation)
while the dynamic veto schemes generally remain efficient or become more efficient (due to harder, final-state charged leptons).
At lower $m_{\tilde{\mu}_R}$ and close to the degenerate limit, final-state leptons are relatively soft.
This leads to $\pTVeto$ and $\HTVeto$ thresholds that are as tight as, if not more stringent than,  the static veto,
thereby eliminating any improvement from relaxing other selection cuts.

Qualitatively, we observe that $H_T$-based vetoes tend to perform better at high masses while $p_T^{j_1}$-based vetoes are better at low masses,
indicating the utility of veto schemes that employ more inclusive measures of the hadronic activity, \textit{e.g.}, $H_T$.
$S_T$-based schemes are competitive. However $\pTVeto = S_T$ is too inclusive for small $m_{\tilde{\mu}_R}$ where the static analysis gives better results. 
The inclusive nature of $S_T$ is particularly useful in the compressed region, where individual lepton momenta are the smallest.
In short, a whole class of dynamic jet vetoes can improve discovery potential of smuon pairs,
but the  difference in performance across the various limits of parameter space suggests that no single combination of hadronic and leptonic activity measures will be ideal in all cases.
The appropriate leptonic measure should be investigated on an analysis-by-analysis basis in order to target specific kinematic regions.

\subsubsection*{Impact of Jet Veto Rapidity Window}

Experimentally, jets can only be reconstructed within the range of the detector, {\it i.e}, with a pseudorapidity $\vert\eta\vert \lesssim 4.5$ for ATLAS and CMS.
In practice though, stringent, static jet vetoes are often only applied within the coverage of the tracker, typically for jets with $\vert\eta\vert\lesssim2.4$.
Extending jet vetoes to the forward region, $2.4 \lesssim \vert \eta \vert \lesssim 4.5$, is avoided, among other reasons,
to help to mitigate the contamination of pile-up activity, including the contribution to low-$p_T$ jets that would otherwise never exceed a veto threshold.
This avoidance, however, is at the cost of an increased dependence on higher order QCD splittings, and hence an increased theoretical uncertainty~\cite{Michel:2018hui}.
On the other hand, it has recently been demonstrated that rapidity-dependent, jet vetoes, in particular one wherein $\pTVeto$ is relaxed for increasing jet pseudorapidity,
can reduce this theoretical uncertainty~\cite{Michel:2018hui}, and are already experimentally viable~\cite{Aaboud:2018xdt}.
Moreover, extending dynamic jet vetoes to the forward region was found to be necessary to ensure a sufficient suppression of SM backgrounds in studies at higher $\sqrt{s}$~\cite{Pascoli:2018heg}.

In this context, we briefly investigate the impact of  a dynamic jet veto when expanding the $\eta$ range of the jet veto-window from $|\eta| < 2.4$ to $|\eta| < 4.5$.
For a widened $\eta$ range, we show in fig.~\ref{fig:sleptonVeto_Mu95Ratio_WideEta}, the signal strength ratio,
\begin{equation}
\mathcal{R}_{X} = \mu_{SS}^{\rm X}(\vert \eta^{\rm Veto}\vert < 2.4) / \mu_{SS}^{X}(\vert \eta^{\rm Veto}\vert < 4.5),
\end{equation}
for (a) the benchmark static jet veto analysis, where $\pTVeto=25\GeV$, and (b) the dynamic analysis, with $\pTVeto = p_T^{\ell_2}$.
As before, a ratio of $\mathcal{R}_{X}>1$ indicates improved sensitivity.
When a static veto is used and the pseudorapidity range increased, the vetoing of jets outside the central region reduces background rates while simultaneously reducing the signal rates,
thereby maintaining a similar signal-to-background efficiency as in the reference analysis.
For the dynamic veto, however,
there is a uniform $\mathcal{O}(5-20)\%$ improvement for most of the parameter space due to slightly higher background rejection coupled with a smaller decrease in signal efficiency.
We anticipate this behavior to hold for all other dynamic veto schemes considered in this analysis.

\subsubsection*{Impact of Jet Vetoes When Lifting The $M_{T2}$ Cut}

As shown in table~\ref{tab:CMSCutflows}, requiring the selection cut $M_{T2} > 90$ GeV  greatly suppresses electroweak diboson and top quark pair production independently of a jet veto.
However, the cut also reduces considerably the signal acceptance when sparticles are mass-degenerate.
Notably, we report that choosing a more aggressive dynamic jet veto can control the top pair background sufficiently in the absence of the $M_{T2}$ cut,
leading to a significant improvement in sensitivity.

We have checked that using $H_T < p_T^{l2}$ as a dynamic veto is stringent enough to control the top pair background when lifting the $M_{T2}$ cut, independently of the signal region.
When relaxing $M_{T2}$, total background rates grow by a factor of 5 for the lowest $\slashed{E}_T$ signal region up to a factor of 1.5 for the highest $\slashed{E}_T$ signal region,
while there is a large, overall increase in signal efficiency.
For the benchmark point $(m_{\tilde\mu_R},m_{\tilde{\chi}_1}) = (750\GeV,~700\GeV)$,
this results in negligible changes in the signal ($S$) over background ($B$) ratio $S/B$ for the two lower $\slashed{E}_T$ signal regions but significant increases in $S/B$ for the two higher $\slashed{E}_T$ signal regions.
 Lifting the $M_{T2}$ cut when using a stringent dynamic veto based on $H_T$ therefore allows for improvements in sensitivity in the compressed region,
 independently of the integrated luminosity, due to the top pair background being sufficiently controlled by the dynamic veto itself.

We find though that the improvement does not hold for all veto schemes considered.
When requiring $p_T^{j1} < p_T^{l2}$ and no $M_{T2}$ restriction,
the top pair background comes to dominate the background rate in the two lower $\slashed{E}_T$ signal regions and increases the rates by factors of $20-30$,
thereby reducing $S/B$, despite the increased signal efficiency.
The two higher $\slashed{E}_T$ signal regions are less affected due to a much smaller the top pair contribution, with only a factor of 2 increase in the total background rate for the highest $\slashed{E}_T$ one.
For $(m_{\tilde{\mu}_R},m_{\tilde{\chi}_1}) = (750\GeV,~700\GeV)$, we see a reduction in $S/B$ in all signal regions, except for the highest $\slashed{E}_T$ one,
suggesting that the simplest incarnations of dynamic jet vetoes are not sufficient in their own right.
This was noted previously in refs.~\cite{Pascoli:2018rsg,Pascoli:2018heg}.

\section{Summary and Conclusion}\label{sec:conclusions}
In summary, we have investigated several measures of leptonic and hadronic activities in the process
\begin{equation}
pp \to \gamma^*/Z^* +X\to \tilde{\mu}_R^+\tilde{\mu}_R^- +X \to \mu^+\mu^- + \slashed{E}_T +X,
\end{equation}
and the associated SM background processes, to explore possible generalizations of dynamic jet vetoes.
Using this information, we have demonstrated that a general class of dynamic jet vetoes can be used to improve the sensitivity of searches for right-handed smuon pair production at the LHC.
The improvement becomes more significant as we probe mass scales further above the EW scales, and in some instances hold even when the final-state particles are soft.
{
Differences between the various processes can be directly attributed to underlying kinematics and QCD radiation patterns,
particularly when radiation amplitude zeros are involved (see sec.~\ref{sec:signalProc_beyondPT}).
} 
Most choices of measures for hadronic and leptonic activities perform better than the CMS-inspired benchmark analysis, 
which features a static jet veto threshold of $\pTVeto=25\GeV$ (see fig.~\ref{fig:sleptonVeto_95CLSensitivity_LHC14_ratio}).
Differences suggest that no single dynamic veto scheme will always be ideal for all parameter space regions 
and rather should be investigated on an analysis-by-analysis basis.
Qualitatively, we find that dynamic jet vetoes using more inclusive measures of the hadronic activity, e.g., $H_T$,
perform best, while the ideal choice of leptonic activity depends on the signal kinematics (see sec.~\ref{sec:results}).
We report that the impact of including MPI/UE and NLO-accurate jet merging, \textit{e.g.,} via the FxFx method, does not appreciably alter this picture; see table~\ref{tab:CMSCutflows}.
The impact of enlarging the jet veto rapidity window, complementarity to other selection cuts,
{and dynamic vetoes built from $\not\!\vec{p}_T$-based observables} were also addressed.

Due to the dynamic nature of these cuts, sensitivity can likely be improved with machine learning techniques such future investigations are encouraged.
We anticipate that our results generalize to other searches for new, heavy, uncolored physics that employ jet vetoes at the LHC,
and push for investigations in this direction.


\section*{Acknowledgements}
Kate Pachal and Fibonacci Tamarit are thanked for discussions over fondue and curry.
Freya Blekman, Dag Gillberg, Ilkka~Helenius, Simon Platzer, and Dieter Zeppenfeld are also thanked for discussions.
This work has been partly supported by French state funds managed by
the Agence Nationale de la Recherche (ANR) in the context of the LABEX
ILP (ANR-11-IDEX-0004-02, ANR-10-LABX-63),
which in particular funds the scholarship of SLW.
This work has received funding from the European Union’s Horizon 2020 research and innovation programme 
as part of the Marie Sklodowska-Curie Innovative Training Network MCnetITN3 (grant agreement no. 722104).
KN is supported by the NWO. KN and SLW acknowledge the generous hospitality of CP3 at UCLouvain.
RR is supported under the UCLouvain MSCA co-fund ``MOVE-IN Louvain,"
the F.R.S.-FNRS ``Excellence of Science'' EOS be.h Project No 30820817,
acknowledges the contribution of the COST Action CA16108,
and acknowledges the hospitality of the LPTHE.


\providecommand{\href}[2]{#2}\begingroup\raggedright

\endgroup

\end{document}